\documentclass[fleqn,usenatbib]{mnras}

\usepackage{newtxtext,newtxmath}

\usepackage[T1]{fontenc}

\usepackage{graphicx}	
\usepackage{amsmath}	

\defcitealias{Parikh2018}{P18}
\defcitealias{Parikh2021}{P21}

\title[Stellar IMF and M/L]{Stellar populations of massive early-type galaxies observed by MUSE}

\author[T. Parikh et al.]{
Taniya Parikh,$^{1,2}$
Roberto Saglia,$^{1,3}$\thanks{E-mail: saglia@mpe.mpg.de}
Jens Thomas,$^{1,3}$
Kianusch Mehrgan,$^{1,3}$
\newauthor{Ralf Bender,$^{1,3}$
Claudia Maraston$^{2}$}
\\
$^{1}$Max-Planck-Institut f\"ur extraterrestrische Physik, Giessenbachstrasse 1, 85748 Garching bei M\"unchen, Germany\\
$^{2}$Institute of Cosmology, University of Portsmouth, Burnaby Road, Portsmouth PO1 3FX, UK\\
$^{3}$Universit\"ats-Sternwarte M\"unchen, Scheinerstrasse 1, D-81679 M\"unchen, Germany\\
}

\date{Accepted XXX. Received YYY; in original form ZZZ}

\pubyear{2024}

\begin{document}
\label{firstpage}
\pagerange{\pageref{firstpage}--\pageref{lastpage}}
\maketitle

\begin{abstract}
Stellar population studies of massive early-type galaxies (ETGs) suggest that the stellar initial mass function may not be universal. In particular, the centres of ETGs seem to contain an excess of low-mass dwarf stars compared to our own Galaxy. Through high resolution MUSE IFU data, we carry out a detailed study of the stellar populations of eight massive ETGs. We use full spectrum fitting to determine ages, element abundances, and IMF slopes for spatially binned spectra. We measure flat gradients in age and [Mg/Fe] ratio, as well as negative gradients in metallicity and [Na/Fe]. We detect IMF gradients in some galaxies, with the centres hosting bottom-heavy IMFs and mass excess factors between 1.5-2.5 compared to a Kroupa IMF. The IMF slope below 0.5~M$_\odot$ varies for our galaxy sample between 1-2.8, with negative radial gradients, while the IMF slope between 0.5-1~M$_\odot$ has a steep value of $\sim$3 with mildly positive gradients for most galaxies. For M87, we find excellent agreement with the dynamical M/L as a function of radius. For the other galaxies, we find systematically higher M/L from stellar populations compared to orbit-based dynamical analysis of the same data. This discrepancy increases with NaI strength, suggesting a combination of calibration issues of this line and correlated uncertainties.\end{abstract}




\section{Introduction}
The stellar initial mass function (IMF) is a crucial ingredient in modelling and interpreting the integrated light of galaxies. It regulates star formation and chemical enrichment throughout the life of a galaxy. The IMF was first discovered to be a power law with a slope of x = 2.35 \citep{Salpeter1955} for stars in the Milky Way with masses \textgreater\ 0.5~$M_\odot$. Later studies revealed a flattening of the power law slope towards low masses \citep{Kroupa2001, Chabrier2003} with a slope of x = 1.3 between 0.1 - 0.5 $M_\odot$ for a Kroupa IMF. In unresolved populations, the IMF can be measured spectroscopically through gravity-sensitive spectral features. The strengths of these features are determined by the ratio of dwarf to giant stars that are present in the stellar population. Alternative methods include constraining the M/L dynamically or through gravitational lensing, and comparing this to the expected M/L for a Kroupa IMF. This ratio of mass-to-light ratios can be thought of as an excess in mass. For massive early-type galaxies, the mass excess was found to increase with velocity dispersion \citep{ThomasJ2011, Cappellari2012, Conroy2012b, Li2017}, and towards the centres of galaxies \citep{MartinNavarro2015a, vanDokkum2017, Parikh2018, LaBarbera2019, Dominguez2019, Gu2022} \citep[see][for a recent review]{Smith2020}.


Investigations into the reason behind IMF variations have revealed correlations with the velocity dispersion, $\alpha$-abundance, metallicity, orbital properties \citep{MartinNavarro2015b, vanDokkum2017, Parikh2018, Poci2022}. Variation has also been observed with kinematics, such that fast rotators (FRs) display bottom-heavy IMFs with steeper radial gradients than slow rotators (SRs) \citep{Bernardi2019}. Combining photometric and kinematic analyses, SRs are cored with a flat central luminosity profile, whereas FRs are cuspy or power law galaxies, with an extra central component \citep{Kormendy2009, Emsellem2011, Lauer2012}. Cored galaxies are thought to be the result of dissipationless (dry) mergers of cuspy galaxies, suggesting that steep IMF gradients become shallower as a result of merging.

The emerging consensus of a varying IMF is still under tension with the results from four strongly lensed galaxies by 2-3$\sigma$ \citep{Newman2017, Collier2018}. Constraints depend on the assumed shape of the IMF, and varying the low-mass cutoff or a non-parametric IMF with free contributions of stars in different mass intervals provides consistent results from lensing and stellar population analysis for all but one of these galaxies. Simulations can also provide insight into this tension as they can be used to study the physical cause of potential IMF variations. \citet{Barber2019} showed using EAGLE simulations that a pressure dependent low-mass or high-mass IMF slope, calibrated to reproduce the mass excess-velocity dispersion relation in \citet{Cappellari2013}, predicts local IMF variations with radius and metallicity as seen in observations. Contrary to this, no departures from a Kroupa IMF have been observed within our own Galaxy, even in the metal-rich bulge which provides environments closer to that of early-type galaxies. Simulations cannot simultaneously explain the IMF variations observed in massive early-type galaxies and the universality of the IMF seen in our own Galaxy \citep{Guszejnov2019}.

Comparisons of dynamical and stellar population mass-to-light ratios on a galaxy-by-galaxy basis play an important role in assessing any possible systematics. \citet{Smith2014} found no correlation between M/L derived by the two methods, however the dynamical results \citep{Cappellari2013} were based on a much larger aperture size than stellar populations \citep{Conroy2012b}. \citet{Lyubenova2016} showed consistency for a subsample of their galaxies when combining constraints from stellar populations and dynamics and when using the same data set over the same apertures. Hence it is crucial to consider radial gradients in the IMF, while noting that these cannot explain cases where the dynamical M/L is higher than the stellar population one. Dynamical studies assuming a constant M/L can affect the mass excess factor by $\sim$0.25~dex \citep{Bernardi2018}. Of the few studies which allow a varying M/L, gradients have always been found \citep{Oldham2018, Collett2018}. Looking at M87, \citet{Sarzi2018} obtain stellar population-based M/L systematically larger by 30\% than those from dynamics \citep{Oldham2018}, although gradients agree well. \citet{Dominguez2019} show agreement between the two methods based on a self-consistent comparison of galaxies from the SDSS-IV MaNGA survey.

In \citet[][hereafter P18]{Parikh2018}, over 300 galaxies from the MaNGA survey were analysed and IMF radial gradients were found for the more massive galaxies in the sample, with Salpeter-like IMFs in the centre, returning to a Kroupa IMF at 1 R$_e$. Although these radial gradients were independent of the stellar population model used, the value of the IMF slope between 0.1 - 0.5 M$_{\odot}$ in the more massive galaxy centres, for example, varied from 2 to 3 depending on the model. Discrepancies also exist depending on the region of the spectrum being fit. \citet{Conroy2012b} show using their models that masking the NaI region strongly influences the resulting IMF slope despite carrying out full spectrum fitting. \citet{MartinNavarro2021} apply a hybrid \textit{full index fitting} approach, that fits each pixel within the bandpass of chosen indices, with TiO indices as the IMF-sensitive features. It would be interesting to see the derived [Ti/Fe] values in order to assess residual degeneracies affecting these indices. \citet{Sarzi2018} derive similar IMF gradients using different combinations of indices (e.g. TiO, NaI, CaT) using \citet{Vazdekis2012} models. In addition to the choice of stellar population models and fitting technique, uncertainties can arise due to the underlying stellar library being incomplete. New stellar libraries such as \citet[][MaStar]{Yan2018} and \citet[][XSL]{Verro2022} will help shed further light on this matter.

\begin{table}
	\centering
	\caption{Galaxy properties for the sample studied in this work. $\sigma$ is the velocity dispersion measured from the central radial bin, the half-light radius R$_e$ is taken from literatures (references given in the text), the PSF FWHM is measured from the broadening of point sources in the object or sky exposures, and redshift z is adopted from HyperLEDA.}
	\label{tab:gal_properties}
	\begin{tabular}{lcccc}
		\hline
		Galaxy & $\sigma$ [km/s] & R$_e$ ["] & PSF ["] & z\\
		\hline
		M87           & 312 & 81.20 & 1.20 & 0.00428\\
		NGC 1407 & 261 & 70.33 & 1.93 & 0.00593\\
		NGC 1332 & 296 & 28.00 & 2.12 & 0.00508\\
		NGC 5419 & 314 & 43.36 & 1.56 & 0.01376\\
		NGC 5328 & 318 & 22.24 & 1.28 & 0.01581\\
		NGC 5516 & 280 & 22.09 & 2.00 & 0.01488\\
		NGC 307 & 183 & 16.45 & 2.10 & 0.01338\\
		NGC 7619 & 294 & 32.00 & 2.00 & 0.01255\\
		\hline
	\end{tabular}
\end{table}

In the present work, we build upon previous studies in three aspects: i) we model individual massive galaxies in detail with increased spatial resolution thanks to MUSE, generating maps of stellar population properties for these galaxies, ii) the sample contains galaxies with different kinematic morphologies to explore in detail trends between slow versus fast rotators and their IMF, and iii) we compare the M/L from stellar populations with the independently derived dynamical M/L from literature to provide strong constraints. We aim to use a galaxy-by-galaxy approach to assess the consistency between these two techniques, and also attempt to investigate variations with kinematics to shed further light on the driver of IMF variations.

In \autoref{sec:Data} we describe our MUSE galaxy sample and binning procedure. \autoref{sec:Results} presents our derived stellar population parameters for all the galaxies including spatial maps. We compare our M/L profiles to dynamical modelling results in \autoref{sec:Discussion} and discuss the implications, with our concluding remarks in \autoref{sec:Conclusion}.



\section{Data}
\label{sec:Data}

\subsection{Galaxy sample}
We use data from the ESO programme 095.B-0624(A) (PI: Jens Thomas) for early-type galaxies observed by the Multi-Unit Spectrograph Explorer (MUSE) mounted on the 8.4m UT4 of the Very Large Telescope (VLT). Of these, NGC 6861 and NGC 4751 are excluded from the present analysis due to poor quality data in the near-infrared, which is crucial for IMF derivation. We also use MUSE observations of M87 \citep{Sarzi2018}, giving a sample of eight galaxies with a median redshift of 0.013. The observations are in the wide-field mode, with a wavelength coverage of 4800 to 9400~\AA\ and a spectral resolution of 1.25~\AA\ (see \citet{Mehrgan2023a} for more details on the observations). NGC 307 and NGC 1332 are fast rotators, while the others are intermediate/slow rotating galaxies.

The properties of these galaxies are listed in \autoref{tab:gal_properties}.  $\sigma$ is the maximum central velocity dispersion as measured in this work, R$_e$ comes from previous publications: M87 from \citet{Gebhardt2009}; NGC 1407, NGC 5328, NGC 5516 and NGC 7619 from \citet{Rusli2013}; NGC 1332 from \citet{Rusli2011}; NGC 307 from \citet{Erwin2018}; NGC 5419 from \citet{Mazzalay2016}. The FWHM of the PSF is measured from the broadening of point sources in the object or sky exposures, while redshifts are adopted from HyperLEDA.

\begin{figure*}
 	\includegraphics[width=\linewidth]{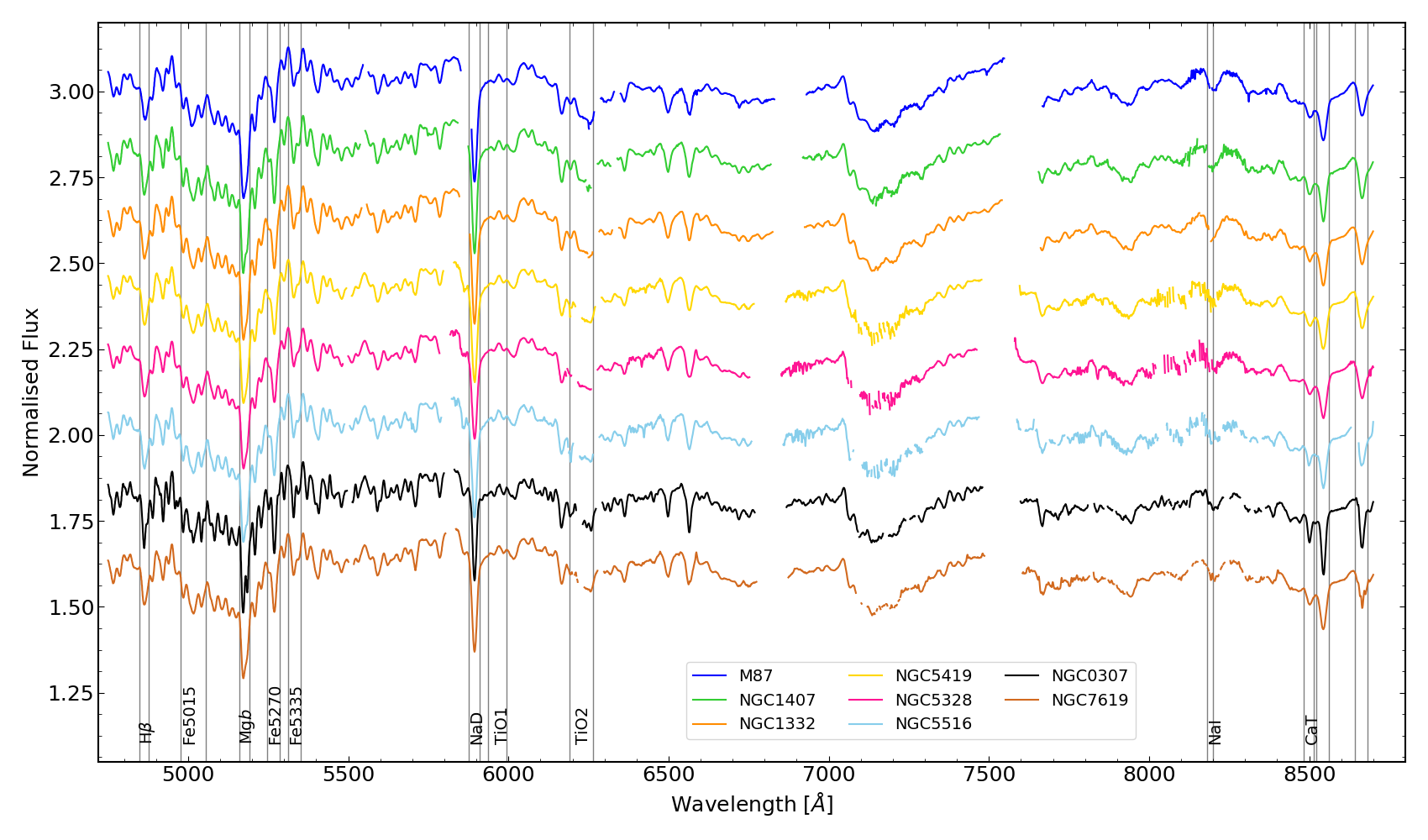}
    \caption{The central radiallly-binned spectrum of each galaxy is shown in the restframe. Key absorption features are labelled and masked regions (due to sky and telluric residuals) are omitted.}
    \label{fig:rad_spectra}
\end{figure*}

Of the galaxies in our sample, M87 \citep{Sarzi2018} and NGC 1407 \citep{vanDokkum2017} have been previously analysed and found to contain steep IMF gradients. The prior is analysed using indices and \citet{Vazdekis2012} models and the latter using full spectrum fitting with Conroy models (same as our approach). However the data for the latter is different. We provide a detailed comparison of our results with these works in the Results section.

\subsection{Spectra}
We proceed to stack galaxy spectra in elliptical annuli for all galaxies, and also in Voronoi bins for galaxies with higher S/N in order to achieve more spatial resolution in these cases. The elliptical annuli are along the semi-major axis and the number of bins are chosen depending on the effective radius coverage within the MUSE FOV, while the Voronoi bins have a target S/N of $\sim400$. Foreground stars and other galaxies in the FOV are masked and individual pixels with S/N \textless 5 are excluded from the binning. There are two sources of error in a binned spectrum: the observational error associated with the data (which includes errors due to sky subtraction), and the error on the mean due to the standard deviation of all spectra within a bin. We combine these in quadrature to derive the error spectrum.

Binned spectra are fit using ppxf \citep{Cappellari2017} to determine stellar kinematics and to subtract emission lines where present. We remove emission associated with H$\beta$, [OIII] $\lambda$4960, $\lambda$5008, [NI] $\lambda$5199, $\lambda$5202, HeI, [OI] $\lambda$6302, $\lambda$6366, H$\alpha$, [NII] $\lambda$6550, $\lambda$6585, [SII] $\lambda$6718, $\lambda$6733. We impose a limit that the [NI] line, just red-ward of Mg$b$, is only subtracted when the emission strength-to-error ratio is greater than 10, since the routine tries to fit something even when no emission is present. This led to [NI] emission being fit only for inner bins, as expected. Spectra are then shifted to rest-frame using the redshift and rotation velocity. For the radial bins, spectra are shifted to rest-frame using the velocity from the Voronoi bins before binning. This greatly improves the spectral quality in the near-infrared for the fast rotators, since the high stellar rotation velocity means that the sky residuals are shifted to slightly different wavelengths and are reduced when averaging spectra together. We mask out three strong sky line regions (5270-5586~\AA, 6295-6315~\AA, 6357-6377~\AA; observed frame wavelengths), two deep telluric regions (6860-6960~\AA, 7580-7700~\AA), and NaD absorption from the Milky Way (5570-5586~\AA).


%

Central binned spectra for all galaxies are shown in \autoref{fig:rad_spectra}. Sky lines and regions with strong telluric residuals are masked and the wavelengths are in the restframe. NGC 5516, NGC 5419, and NGC 5328 are all in particular affected in the NaI region near 8200~\AA, due to their redshifts. For M87, the central 3" and the jet of material from the AGN are masked, since the continuum is contaminated by synchrotron emission in these regions.

\subsection{Stellar population models}
For these massive galaxies we require stellar population models at high metallicities, along with a large grid of IMFs and element abundances. For this purpose the models of \citet{Conroy2018} are presently the most suitable, based on the stellar library described in \citet{Villaume2017} and theoretical response functions for various elements. The stars come from MILES and extended IRTF libraries providing super-solar metallicity coverage, topped up with dwarf stars from \citet{Mann2015} to cover the cool dwarf regime.

\begin{figure*}
	\includegraphics[width=\linewidth]{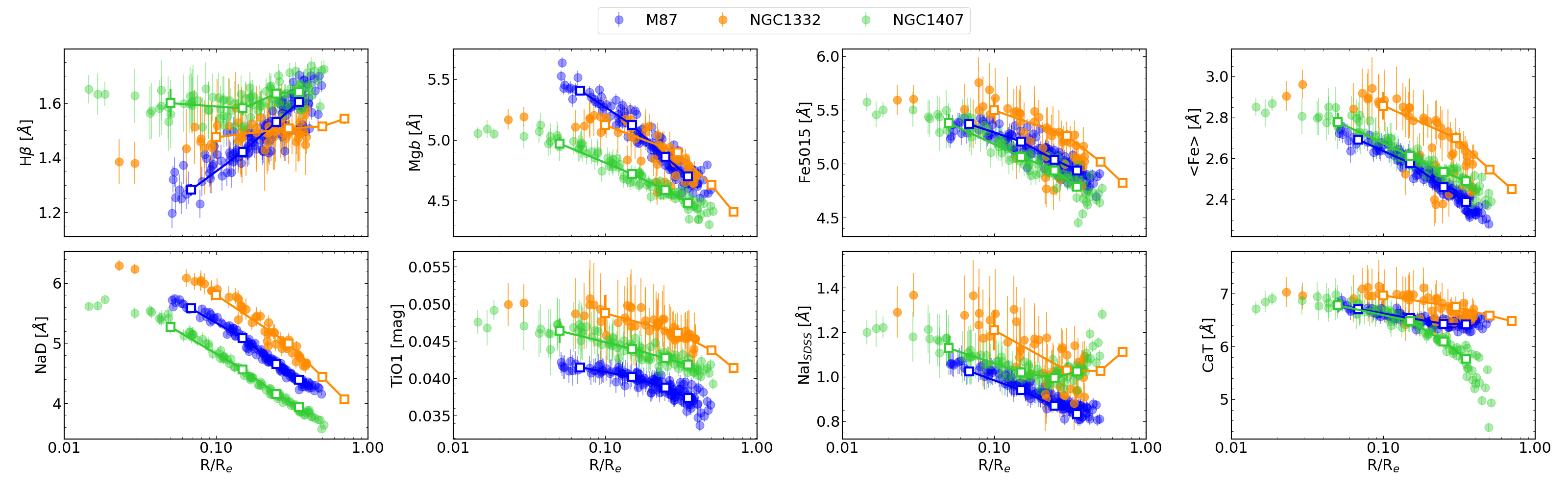}
    \caption{Key absorption line strengths are shown as a function of radius for the three galaxies studied in greater detail. Filled symbols are measurements on the Voronoi binned spectra, while open symbols are measured on the elliptical annuli. The two binning schemes provide consistent results. All indices show negative radial gradients except for H$\beta$, which is flat or increases mildly with radius. NaD shows a particularly steep rise towards the centre.}
    \label{fig:indices_vor}
\end{figure*}

\begin{figure*}
	\includegraphics[width=\linewidth]{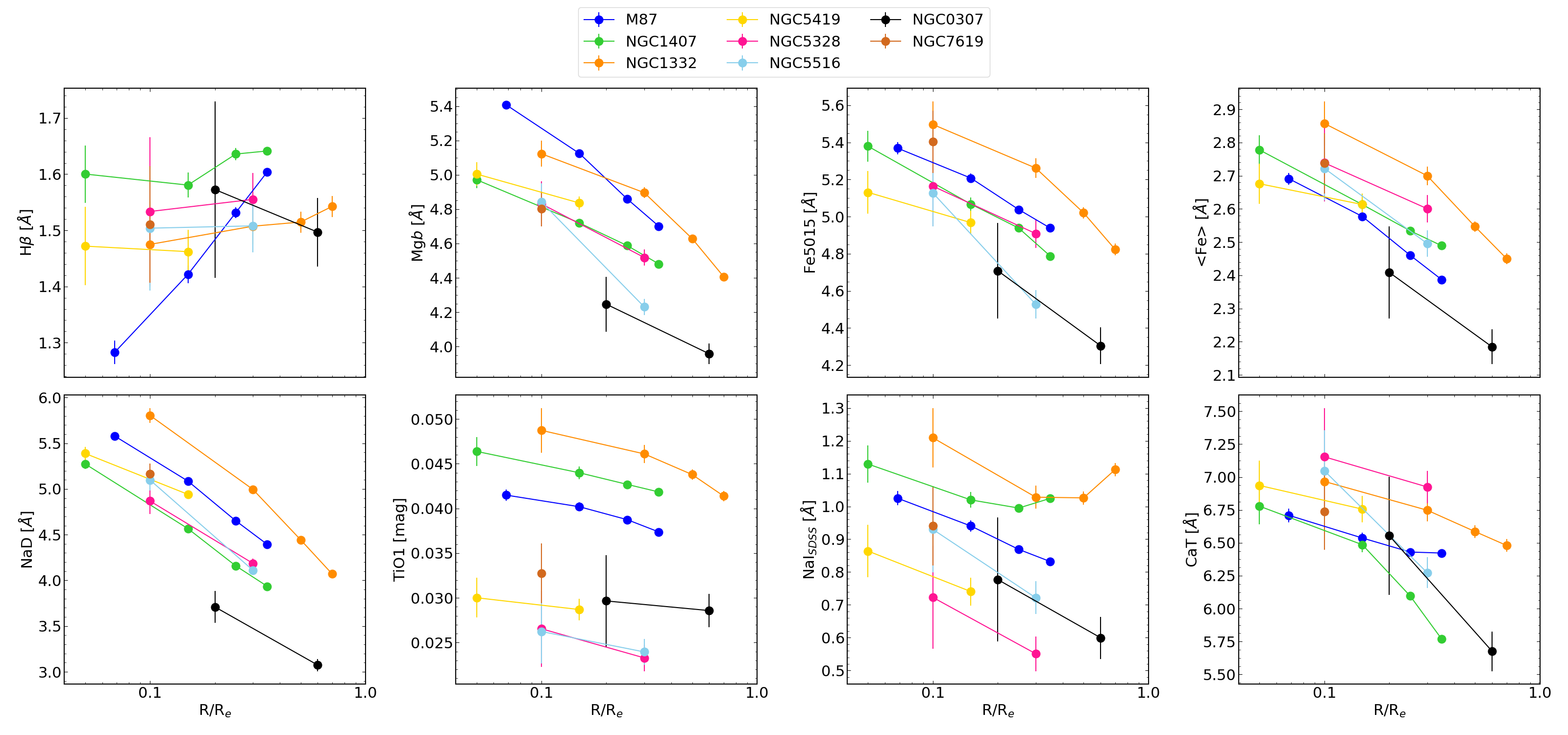}
    \caption{Same as \autoref{fig:indices_vor} measured on the elliptical annuli bins for the full sample of galaxies. All galaxies have high velocity dispersions; NGC 307 which has lower $\sigma$ is shown in black. H$\beta$ falling towards the centre for M87 is likely caused by residual emission. M87 also has very strong Mg$b$ absorption, and relatively weaker <Fe>. NGC 1332, NGC 1407 and M87 have the strongest IMF-sensitive indices TiO1 and NaI. Generally galaxies with stronger TiO and NaI display weaker CaT, and vice versa (with the exception of NGC 1332).}
    \label{fig:indices_rad}
\end{figure*}

Fitting the full spectrum or fitting absorption strengths are complementary approaches for deriving stellar population parameters. While full spectrum fitting makes use of all available information and leads to smaller errors compared to fitting indices for spectra of the same S/N \citep{Conroy2018}, indices exploit key information encoded within narrow wavelength regions, avoiding regions of the spectrum which are noisy or not well understood. We measure absorption strengths using equations 2 and 3 from \citet{Worthey1994}, with the optical index definitions from \citet{Trager1998}, NaI$_{SDSS}$ from \citet{LaBarbera2013}, and CaT from \citet{Conroy2012a}. The measurements are corrected to a common resolution using correction factors calculated from the \citet{Maraston2020} stellar population models. To do this, we measure the indices on the models at two dispersions: the $\sigma$ of the galaxy bin as derived from ppxf, and at 250km/s. We use the ratio between these as a correction factor to obtain indices at 250km/s resolution.

\begin{figure*}
  \includegraphics[width=\linewidth]{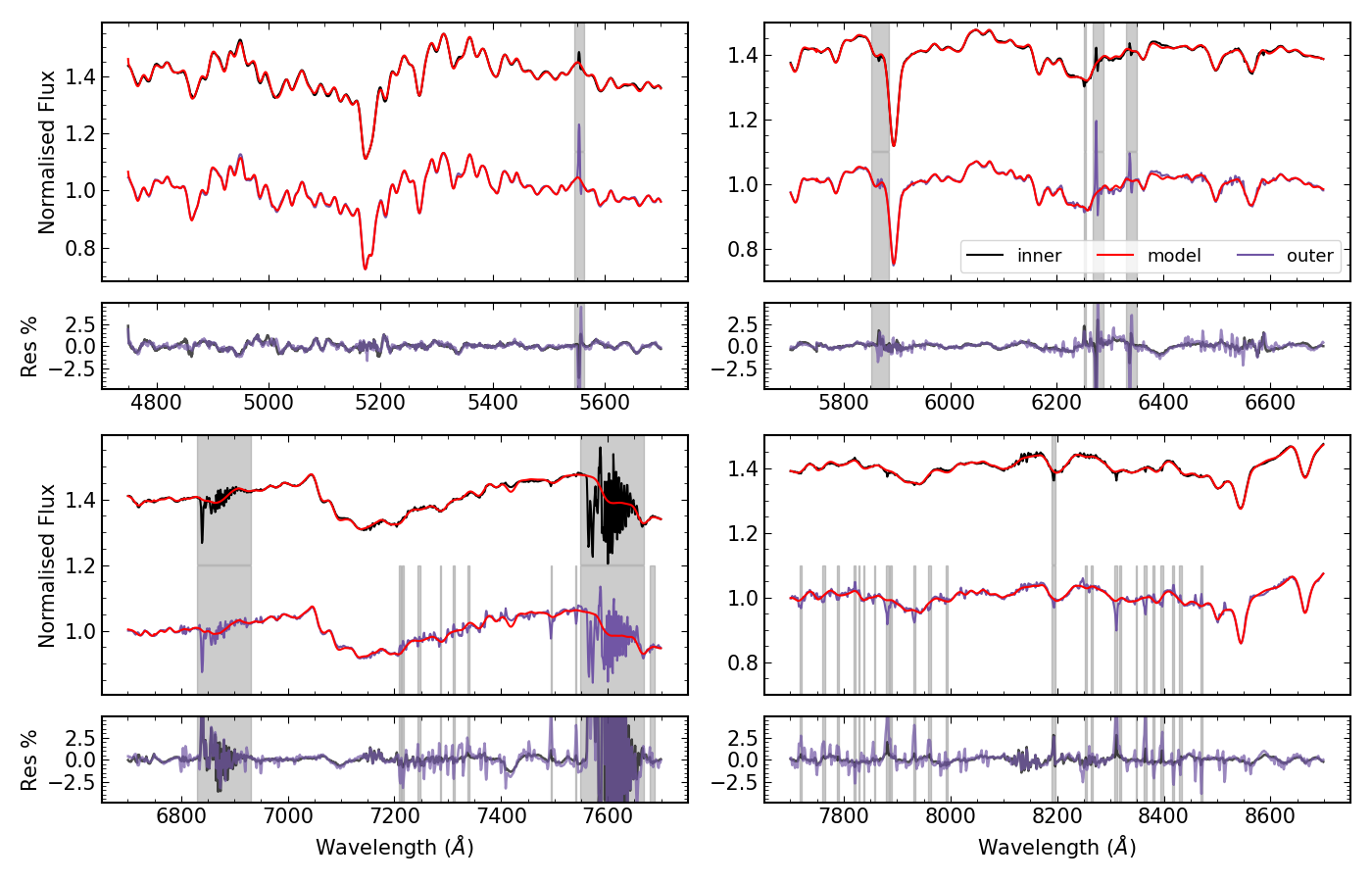}
  \caption{The best-fitting stellar population models (red) to binned M87 spectra are shown for the central bin (black) and an outer bin at 0.5~R$_e$ (purple). The spectra have been offset for clarity. Masked regions due to strong sky and telluric residuals are highlighted in grey. The residuals are generally below 0.5\% in the centre and $\sim$ 1\% at larger radius.}
\label{fig:radial_fit_M87}
\end{figure*}

We make use of the full spectrum fitting code ALF \citep{Conroy2012a, Conroy2018} which explores up to 42 parameters with MCMC. Compared to the \citet{vanDokkum2017} setup, we have the following differences: we allow the higher order Gauss-Hermite moments, h3 and h4, to vary; the element abundances Cu/H and Eu/H are also kept free; the low-mass cutoff of the IMF is fixed to 0.08. Unfortunately we are unable to vary the latter since these models are not publicly available. We tested keeping the other parameters fixed and found no significant effect on our results. We show the Gauss-Hermite parameters in \autoref{sec:app_h3h4} and plan to implement a fully non-parametric LOSVD as an input to the full spectrum fitting code in the future, motivated by the results of \citet{Mehrgan2023a}.

The IMF slope can be varied in two regions: between 0.08 - 0.5 M$_\odot$ and 0.5 - 1 M$_\odot$, referred to as X1 and X2, respectively. We split the fit into four regions: 4750-5750\AA, 5750-6750\AA, 6750-7750\AA, 7750-8700\AA, such that within each region the continuum is normalised via a polynomial with one order per 100\AA. Thus the fits begin close to the blue limit for MUSE, and end just after the CaT triplet, since spectra red-ward of this are too noisy to provide any useful information. For these models and code, the metallicity is quoted as [Fe/H]. We convert this to total metallicity [Z/H] using Equation 13 in \citet{Johansson2012}.

\subsection{Residual telluric absorption}
The telluric correction, calculated using a standard star as described in \citet{Mehrgan2023a}, has an associated statistical error. We carried out an exercise to estimate errors due to the uncertainties in the telluric correction. We produced five mocks for NGC 1407 by perturbing the spectrum using this error. Calculating the root mean squared difference for all parameters between each mock and the results presented in this work showed that the effect on the IMF slope X1, and M/L, was significantly large. [Z/H] was also affected, but for all other parameters the rms was comparable to or smaller than the errors derived by ALF. We add this uncertainty in quadrature to the error estimates from ALF. 

\section{Results}
\label{sec:Results}
In this section we present the results of our analysis, starting with the absorption strengths, followed by the derived stellar population parameters. For the absorption strengths, we first show the Voronoi binned data and overlay the radial data for the three galaxies with detailed analyses, followed by the radial results for all eight galaxies. This allows us to demonstrate consistency between the two binning techniques, i.e. Voronoi and elliptical annuli, and present detailed results where available.

\subsection{Index measurements}
\label{sec:index_measurements}
\autoref{fig:indices_vor} shows measurements of key indices as a function of radius for the three galaxies studied in greater detail. The measurements have been corrected to 100km/s (the resolution of the stellar population models) and 1-$\sigma$ errors are shown. Filled symbols represent measurements from the Voronoi bins, and open symbols are results from the elliptical annuli. These binning methods appear consistent with each other, providing confidence in the other galaxies with just radial binning.

\begin{figure*}
  \includegraphics[width=.9\linewidth]{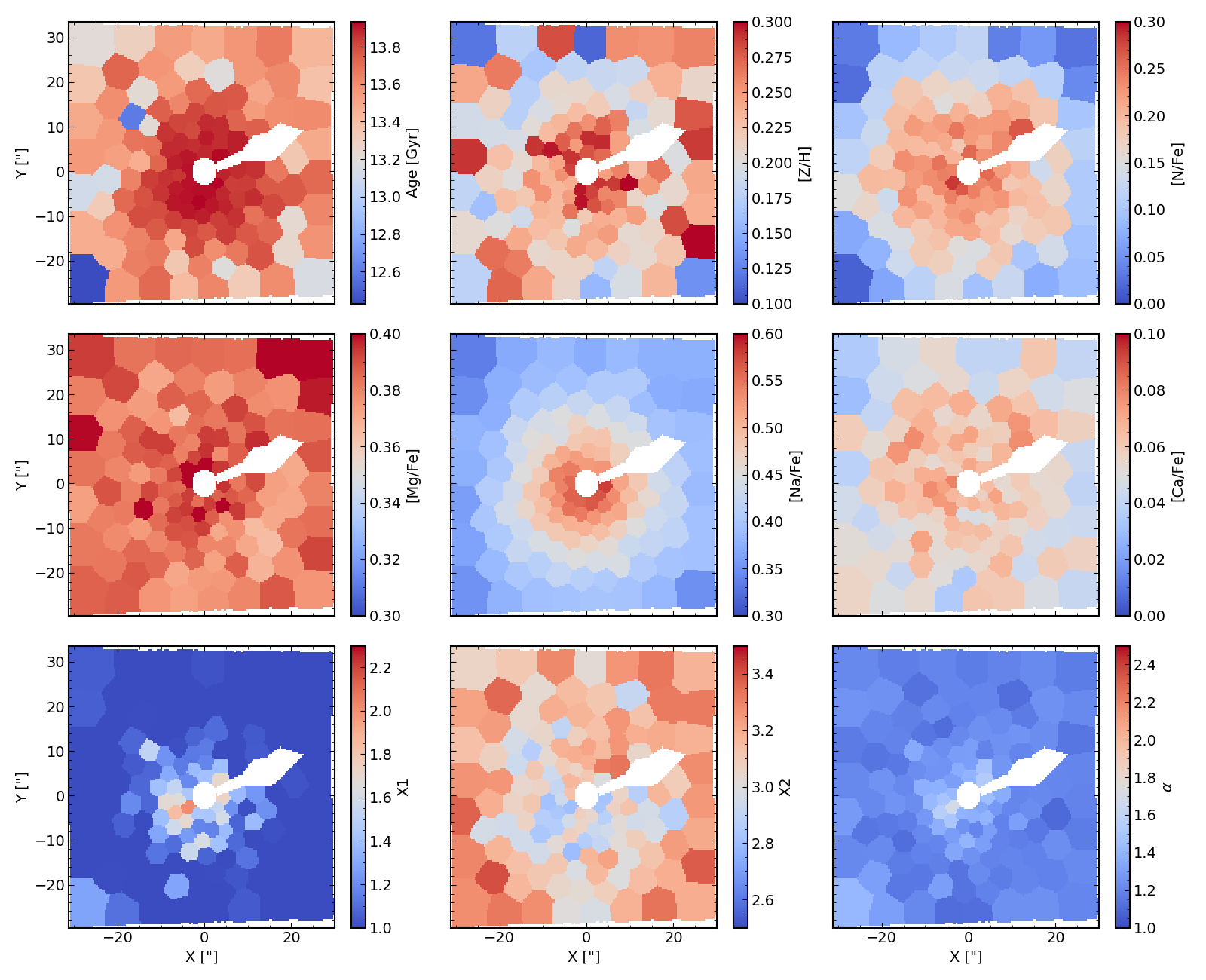}
\caption{Maps of stellar population parameters derived from full spectrum fitting are shown for M87. The FOV covers the galaxy out to $\sim40"$, which is 0.5~R$_e$. The galaxy displays an old stellar population with a negative metallicity gradient. Uniform abundance patterns are seen for Mg and Ca, while N and Na display negative gradients. The IMF slope X1 is found to have a radial gradient from 1.8 in the galaxy centre to below 1 in the outer parts; X2 has larger values $\sim$3, leading to a mass excess of $\sim$1.5 compared to Kroupa.}
\label{fig:M87_maps}
\end{figure*}

Massive galaxies are known to have stronger metallic indices and weaker H$\beta$ \citep{Bender1992, Trager1998}, as we also see here. All indices display negative radial gradients except for H$\beta$, which is either flat or positive (in the case of M87). This is an age-sensitive index and a positive gradient indicates slightly younger stellar populations in galaxy outskirts. M87 shows relatively strong Mg$b$ absorption and weaker Fe indices, which is important because the combination of these indices sets [Mg/Fe] and provides an indication of star-formation timescales. Looking at NaD, all galaxies show strong absorption, which steepens considerably towards the centres, this is linked to highly enhanced [Na/Fe] found in centres of massive early-type galaxies. These intrinsic differences in the spectra indicate the presence of different stellar populations.

The IMF sensitive NaI index is also very strong in galaxy centres with values between 1 - 1.4~\AA. NGC 1332 and NGC 1407 appear to rise at larger radii, which is likely as a result of systematics due to residual sky and tellurics. Note that the annuli for NGC 1332 cover larger radii than the Voronoi bins because shifting spectra before binning helps reduce sky and telluric residuals. High rotation velocities in this galaxy lead to these residuals being shifted to different wavelengths, which are therefore reduced when averaging. The rising NaI at large radii in both the NGC 1407 Voronoi bins and the NGC 1332 radial bins suggests that the data might not entirely be reliable in the NIR. These pixels are masked during the full spectrum fitting and hence do not affect the derived parameters.

Next, \autoref{fig:indices_rad} shows the same equivalent widths for radial bins for the full sample of galaxies. For NGC 7619, we are only able to analyse a single binned spectrum out to 0.2~R$_e$. The clear positive H$\beta$ gradient in M87 is seen here, but central values could be underestimated due to poorly subtracted emission. Since this is the only galaxy with emission, the positive gradient is unlikely to be physical and suggests that the emission lines in M87 are not well modelled. We test our full spectrum fitting results when H$\beta$ is masked and find little to no effect on the age and other parameters, suggesting that the fitting is robust to emission residuals. Looking at the IMF sensitive indices it is important to note that NGC 1332, NGC 1407 and M87 present both strong NaI and strong TiO absorption. NGC 307 has the lowest dispersion which leads to weaker index strengths e.g. Mg$b$, <Fe>, NaD relative to the other galaxies, but note that it has relatively high NaI line-strengths with central values reaching 0.8~\AA.

There appears to be an anti-correlation between TiO1 and NaI index strengths with CaT, while noting that NGC 1332 does not follow this pattern. It is also interesting to see that while e.g. NaD shows larger local (within the galaxy) gradients than across the sample of galaxies, TiO1 shows much larger global (across galaxies) variation compared to its mild local gradients.

\begin{figure*}
 	\includegraphics[width=\linewidth]{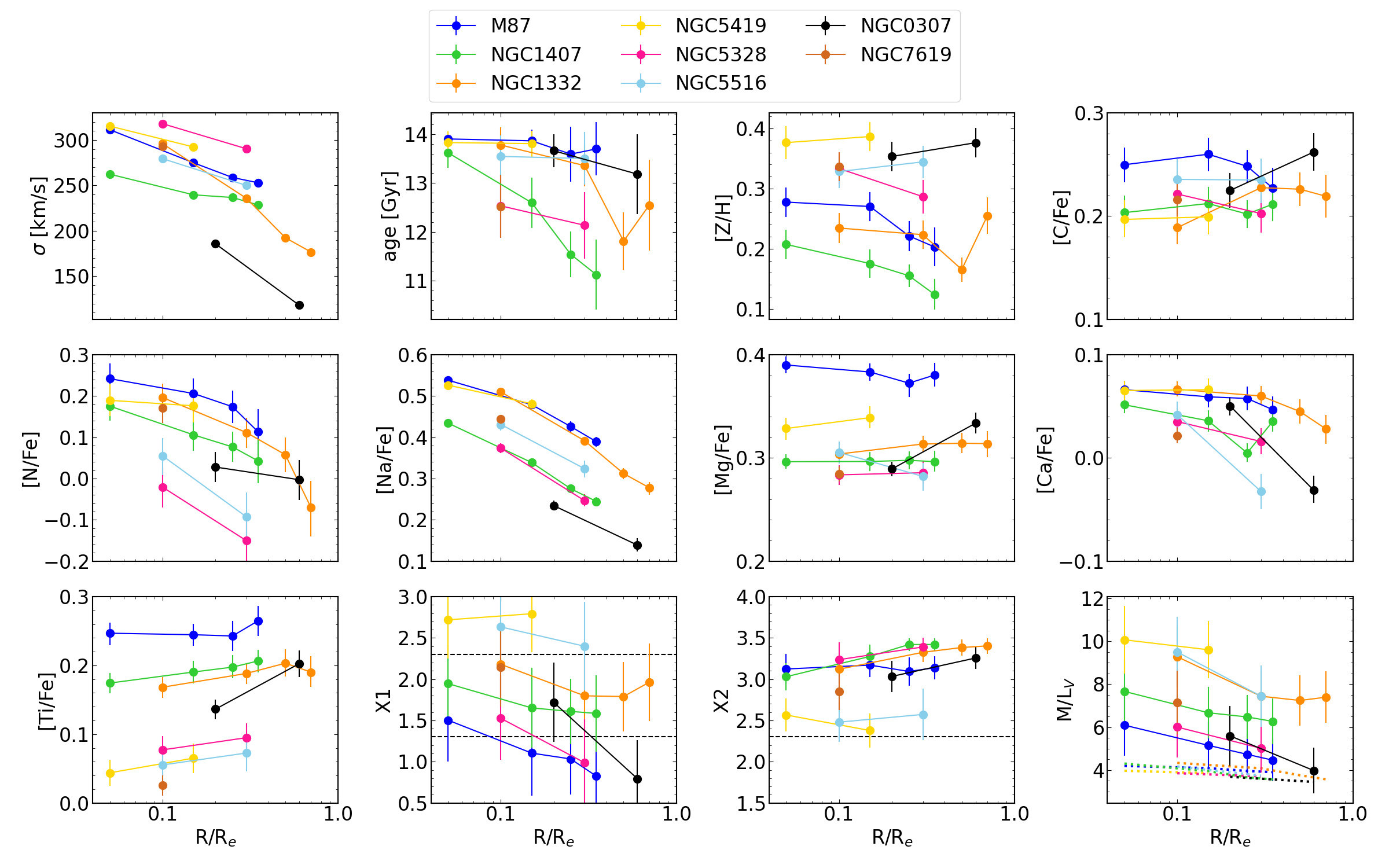}
    \caption{Stellar population parameters as a function of radius for the entire sample of galaxies. There are clear differences between the galaxies: M87 is the most Mg-rich (as well as having high abundances for the other elements). Strong negative gradients in [Na/Fe] are found for all galaxies and in the IMF slope X1 for some galaxies. The dashed lines in the IMF slope X1 panel correspond to Salpeter and Kroupa IMFs at 2.3 and 1.3 respectively, and the dashed line for X2 at 2.3 represents both. Finally, the coloured dashed lines in the final panel show M/L when the IMF is fixed to Kroupa.}
    \label{fig:alfpars_rad}
\end{figure*}

\subsection{Full spectrum fitting}
The fits from the full spectrum fitting code are shown for M87 in \autoref{fig:radial_fit_M87} for a radial bin from the centre within 0.1~R$_e$ and near 0.5~R$_e$. These data are shown in black and purple respectively, and the best-fitting model in red. Grey boxes highlight regions masked due to sky and telluric residuals. The fits to the data are high quality, well within the error at most wavelengths with residuals \textless 0.5\%.

The spectrum from larger radii shows issues with sky and telluric residuals in the near infrared. We mask additional regions due to sky and telluric residuals using a threshold when the sky or telluric signal compared to the flux is \textgreater $5\%$. This limit provided a means to mask a majority of the large spikes seen in the spectra while also ensuring not too much of the data is excluded. Naturally there are more masked pixels in outer bins as the signal from the galaxy decreases. We tested applying the maximum mask from the outermost bin to all spectra and the results were not affected. Prior to this, we discard some outer radial bins with too many masked pixels in the NIR, for all galaxies except M87, NGC 1407 and NGC 1332. Appendix~\ref{sec:app_fits} shows the fits to the other galaxies for the central and outer radial bins.



\subsection{Stellar population parameters}
\label{sec:sp_parameters}

\autoref{fig:M87_maps} shows spatial maps of the stellar population parameters for M87 in the entire MUSE FOV, which goes out to 0.5~R$_e$. The age is uniformly old between 13 - 14 Gyrs (note that the full spectrum fitting code has a prior on the age parameter, restricting its value to younger than the age of the universe), despite the positive H$\beta$ index gradient, which is likely affected by emission. Hence full spectrum fitting circumvents this issue. There are signs of slightly younger stellar populations at larger radii, however this needs to be studied out to larger distances than probed in this work. There is a metallicity gradient with twice-solar metallicities in the centre, falling to 0.1~dex at 0.5~R$_e$. These age and metallicity profiles are as expected for early-type galaxies.

Moving on to the element abundances, [Mg/Fe] and [Ca/Fe] are uniform with values of 0.40, 0.05~dex respectively. These trends are consistent with literature such that Mg is enhanced to twice-solar values, while Ca is relatively under-abundant. Mild gradients are seen in N: [N/Fe] has central values of 0.25~dex decreasing to $\sim$0.1~dex moving outwards. [Na/Fe] is the most enhanced with central values of 0.6~dex and displays a steep gradient, falling to 0.4~dex/R$_e$ near the outskirts. The lower mass IMF slope, X1 has a shallow negative gradient, while X2 has large values of $\sim$3 at all radii. Lastly, the mass excess factor, $\alpha_{Kroupa}$ has a mild gradient from 1.5 in the centre to 1 at larger radii. The mass excess factor is the best-fit M/L divided by the expected M/L for a Kroupa IMF.

\begin{figure}
 	\includegraphics[width=\linewidth]{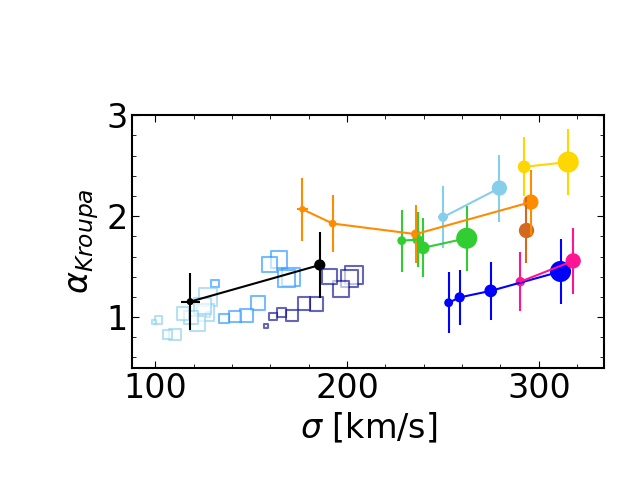}
    \caption{The mass excess factor compared to a Kroupa IMF is shown as a function of the velocity dispersion. Radial bins are shown for each galaxy connected by lines. Also shown for comparison are the results based on MaNGA data from \citetalias{Parikh2018} for three mass bins with darker colours representing more massive galaxies. For both data-sets, symbol size decreases as radius increases. The present results extend the MaNGA analysis to more massive objects and show that the stellar population models predict an increasingly larger mass excess with increasing $\sigma$. M87 and NGC 5328, despite having much larger $\sigma$ than NGC 307, have a similar mass excess to it.}
    \label{fig:alphasigma}
\end{figure}

The parameters from \autoref{fig:M87_maps} plus dispersion and additional elements C, Ti, are now shown as a function of radius in \autoref{fig:alfpars_rad} for the elliptical annuli bins of the full sample of galaxies. All galaxies display old stellar populations with slightly younger ages at large radii. Metallicity gradients, where present, are negative; NGC 5419, NGC 5516, and NGC 307 appear to be very metal-rich \textgreater 0.3~dex out to the radius we are able to study.

Interesting results are seen for different element abundances. [Mg/Fe] values are tightly constrained around 0.3~dex for all galaxies at all radii. M87 however has higher values of 0.4~dex and combined with old ages, this means that these stellar populations formed very early on and quickly. [C/Fe] values are contained within 0.20-0.25~dex, with no gradients, while [Ti/Fe] shows a larger spread of values between solar and twice-solar, again with no significant gradients. [N/Fe] has similar values with negative gradients. [Ca/Fe] values are between solar and 0.1~dex, as previously found for massive early-type galaxies \citep{Saglia2002, Cenarro2003}. Lastly, [Na/Fe] shows steep radial gradients in all galaxies. These profiles reflect the NaD index strengths measured in \autoref{fig:indices_rad}.

We are able to compare our results for M87 and NGC 1407 with abundance profiles from literature for some elements. Our results agree with \citet{Sarzi2018} for [Z/H] and [$\alpha$/Fe] (comparing to our [Mg/Fe]) in the centre but owing to their steeper gradients, their outer values are smaller at -0.1~dex and 0.3~dex (compared to our 0.1~dex and 0.4~dex). They also have steeper [Na/Fe] gradients, such that they find a central maximum values of 0.7~dex (compared to our 0.55~dex), while we agree in the outer regions. We find qualitatively consistent results with \citet{vanDokkum2017} for C, Ca, Mg and Na. A detailed inspection reveals minor differences of the order of 0.05~dex (we find higher central C and Ca and lower Na). The largest difference is that our [Fe/H] are offset to lower values. While they report central values of 0.1~dex, falling to -0.25~dex at our radial limit, we find -0.05~dex at the centre with a mild decrease to -0.1~dex. These differences are likely due to a combination of the wavelength ranges used in all the studies and the different stellar population models in \citet{Sarzi2018}. Note that although both works are based on the same observations, differences can arise due to the method of sky and telluric correction in the data reduction pipelines.

\citet{Parikh2021} found negative gradients for [Na/Fe], while the other five elements were uniform with radius for early-type galaxies. All elements showed clear trends with mass or velocity dispersion. We include a detailed discussion of our element abundances in \autoref{sec:disc_abundances} and compare with literature values, particularly \citet{Parikh2021}.

The final three panels show the IMF slopes X1, X2 and the best-fitting M/L ratio in the V band. X2, the IMF slope between 0.5-1.0~$M_{\odot}$ has a value of $\sim$~2.3 for both Kroupa and Salpeter IMFs, as indicated by the dashed line. In the panel for X1, the slope between 0.08-0.5~$M_{\odot}$, dashed lines indicate Kroupa and Salpeter IMFs. X1 shows much more variation across galaxy centres and within galaxies, and mostly negative gradients. X2 has values of $\sim3$ at all radii and appears to be more tightly constrained with smaller error-bars. \citet{vanDokkum2017} also find negative radial gradients in X1, with values of $\sim$3.0 in the centre to 1.5 at larger radii. Instead for X2, they find values of $\sim$ 2.1, with a mild increase to 2.4 for the outermost bins. 

X1 and X2 appear to be anti-correlated and NGC 5419 is the most extreme example. An anti-correlation between the IMF slopes has also been found by \citet{Feldmeier-Krause2020}, for the galaxy NGC 3923. Note that the parameters derived by \citet{vanDokkum2017} do not show this. Therefore all galaxies have an enhanced fraction of dwarfs between 0.5 - 1~$M_{\odot}$; M87, NGC 5328 and NGC 307 have gradients in X1 ranging from between Kroupa and Salpeter in the centre to below Kroupa at larger radii, NGC 5419 and NGC 7619 display super-Salpeter IMFs, with the other galaxies in between. This is reflected in the  stellar M/L in the V band. Also shown as dotted lines in the last panel is the M/L ratio for a Kroupa IMF. Hence, gradients in the dotted lines are due to only the age or metallicity. It is clear that the largest impact on the M/L value and slope is due to the IMF.

Next, \autoref{fig:alphasigma} shows the mass excess factor as a function of the velocity dispersion. The galaxies from this work are shown as before and the symbol size indicates radius such that larger circles correspond to galaxy centres. Also shown are the results from \citetalias{Parikh2018} as open squares. These are three mass bins spanning 9.9 - 10.8 log M/M$_\odot$, where the higher mass (higher velocity dispersion) bins show a gradient between Salpeter and Kroupa, and the lowest mass bin displays a Kroupa IMF at all radii. The present work extends this to higher masses showing that more massive galaxies have even steeper IMFs. In terms of the gradients, we see a variety such that some galaxies show a gradient while others are bottom-heavy out to our radial coverage. Hence we find an $\alpha$-$\sigma$ correlation but see that individual galaxies show scatter in this relation.

The mass excess factor is between 1 to 2.5 for all galaxy centres. NGC 307 displays the same IMF slope and gradient as slow rotators of much larger velocity dispersions (M87 and NGC 5328). NGC 307 has comparable dispersion to the MaNGA results from \citetalias{Parikh2018}, and has similar or slightly larger IMF than these. It is important to note that these analyses differ in method (full spectrum fitting versus indices) as well as stellar population models.

\begin{figure*}
 	\includegraphics[width=\linewidth]{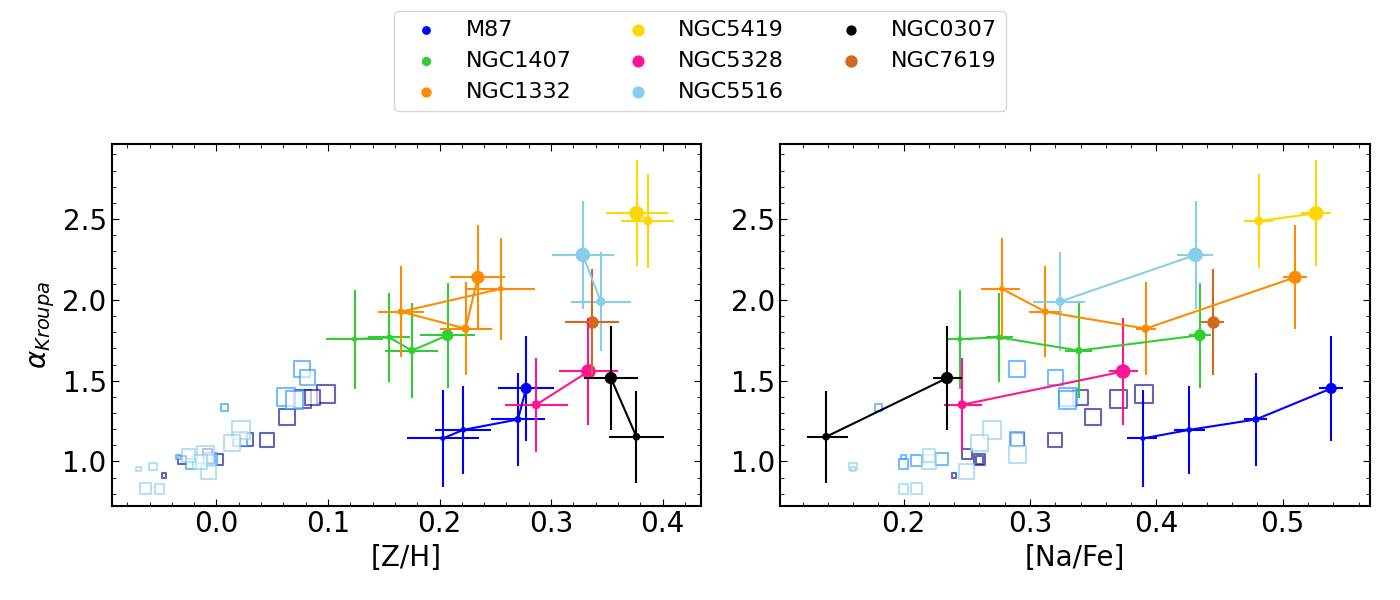}
    \caption{Correlations of the mass excess factor with metallicity (left) and with Na abundance (right) are shown. Symbol colours and sizes as before from \autoref{fig:alphasigma}. Although results for the individual galaxies lead to more scatter, the correlations found from the MaNGA data appear to hold for more massive galaxies. M87 has very high [Na/Fe] and low [Z/H] for its mass excess.}
    \label{fig:correlations}
\end{figure*}

We also set out to explore IMF trends with kinematic morphology. Of the two fast rotating galaxies in our sample, NGC 1332 shows a bottom-heavy IMF and large M/L. The other fast-rotator, NGC 307, also shows a steeper IMF relative to its low dispersion. Yet similar variations are seen amongst the slow rotators. Hence, we see some hints that cuspy galaxies might have more bottom-heavy IMFs than cored galaxies on average, however this can only be confirmed with a larger galaxy sample. \citet{Bernardi2019} find that fast rotators have more bottom-heavy IMFs and steeper gradients than slow rotators for a large sample of galaxies from the MaNGA survey.


We explore correlations between stellar population parameters in \autoref{fig:correlations}. The mass excess is plotted against the metallicity and Na abundance. Decreasing symbol size indicates increasing radius, and results from \citetalias{Parikh2018} are shown as square symbols, as before. There is scatter compared to the results of the MaNGA data in average mass bins, but the correlation appears to extend when studying higher mass galaxies. M87, NGC 5328 and NGC 307 appear to have less mass excess for their metallicities. M87 also has low mass excess compared to its high [Na/Fe], but for the other two galaxies the lower Na abundance could explain the deviation from the $\alpha$-[Z/H] relation.

\section{Discussion}
\label{sec:Discussion}

\subsection{Comparison of element abundances}
\label{sec:disc_abundances}
The accuracy of abundances directly affects the IMF slope. For example, if [Na/Fe] is under- or over-estimated, the NaI index would receive less or more contribution from Na abundance and consequently the IMF would be over- or under-estimated.

\begin{figure*}
  \includegraphics[width=.33\linewidth]{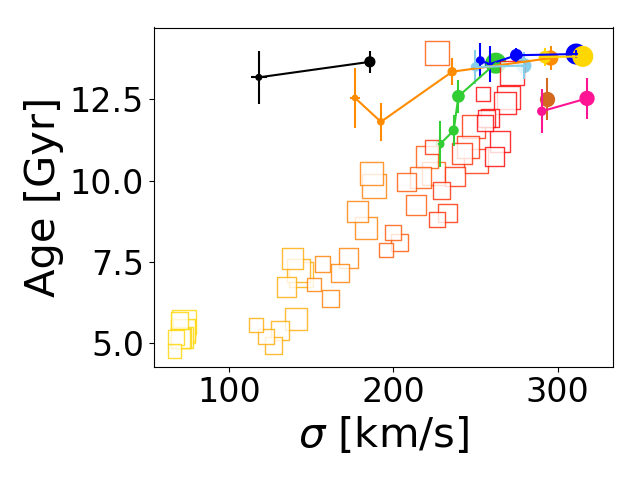}
  \includegraphics[width=.33\linewidth]{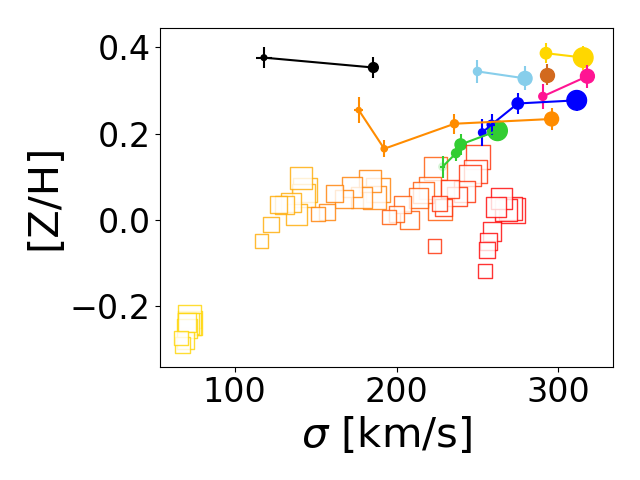}
  \includegraphics[width=.33\linewidth]{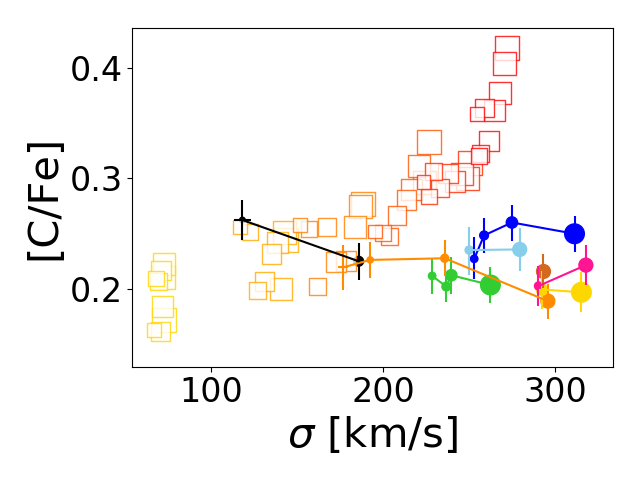}
  \includegraphics[width=.33\linewidth]{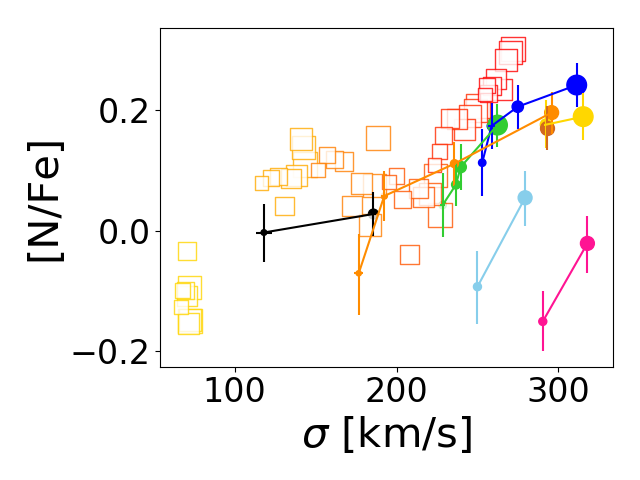}
    \includegraphics[width=.33\linewidth]{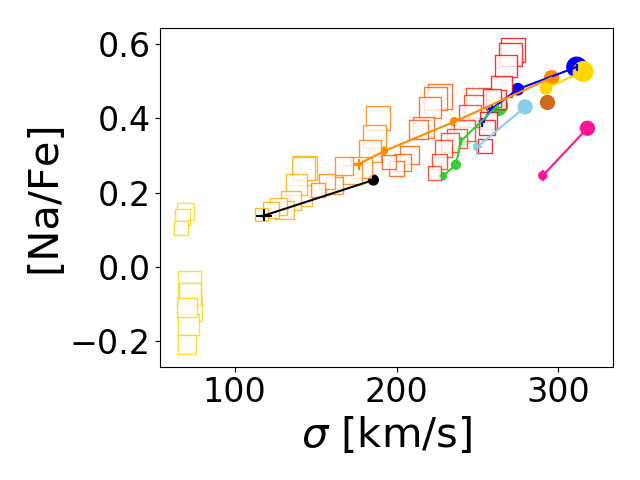}
    \includegraphics[width=.33\linewidth]{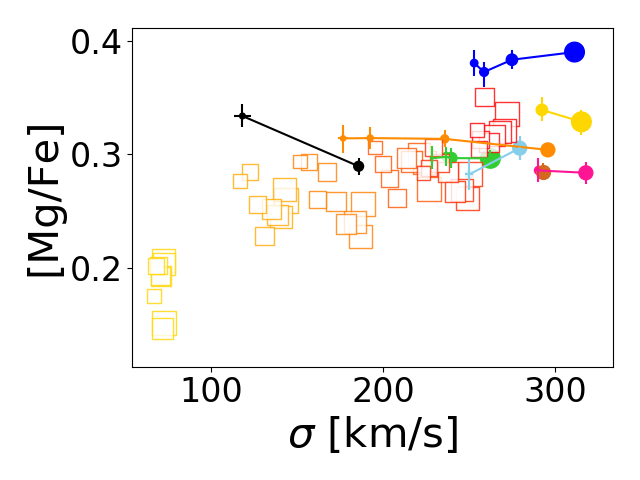}
    \includegraphics[width=.33\linewidth]{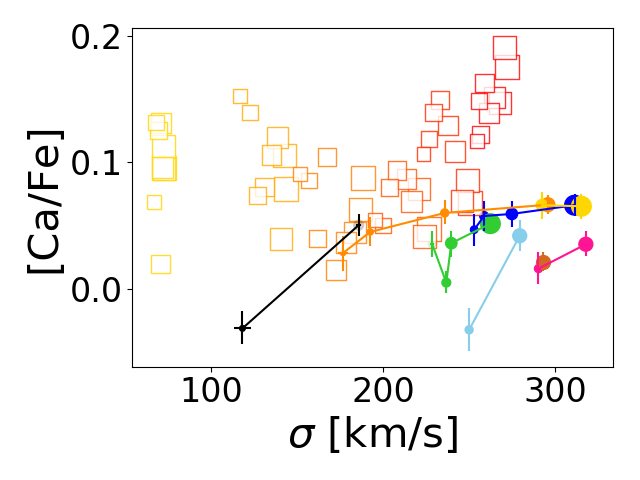}
    \includegraphics[width=.33\linewidth]{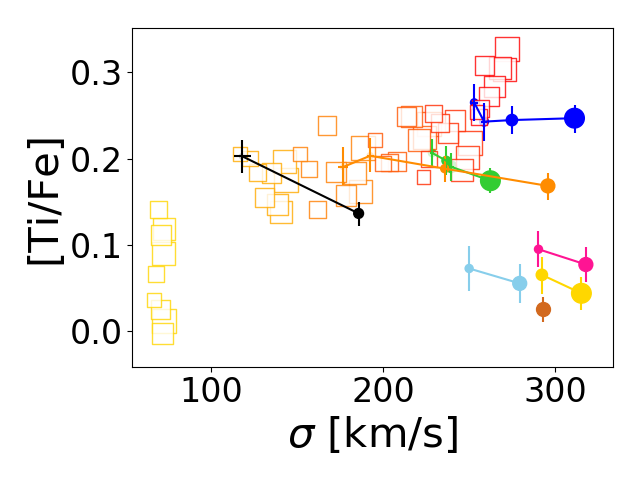}
    \includegraphics[width=.33\linewidth]{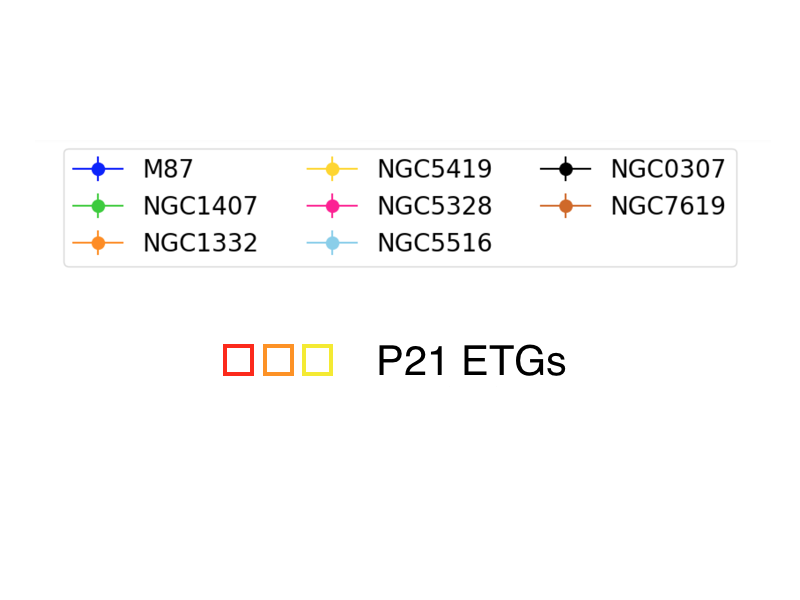}
\caption{Stellar population parameters are shown against the local velocity dispersion for all galaxies. Square symbols show the results from \citetalias{Parikh2021} based on MaNGA data for early-type galaxies in six mass bins. Decreasing symbol size represents increasing radius, as before. The present galaxies correspond to the highest mass bin (red) from MaNGA or are even more massive. The galaxies in the present analysis are older and more metal-rich, but have shallower gradients. There are differences in the abundances of C, Ca, Ti possibly due to the treatment of these elements in the respective stellar population models and the available wavelength range.}
\label{fig:MUSE_P21}
\end{figure*}

We compare our parameters with the results for early-type galaxies from \citetalias{Parikh2021} in \autoref{fig:MUSE_P21}. These are based on six mass bins between 8.8 - 11.3 log(M/M$_{\odot}$) and plot the results out to 1~R$_e$ here. These are shown as open squares, where galaxy mass increases from yellow to red shades. The present analysis is shown for individual galaxies as coloured symbols connected by lines, as in the previous plots. All but NGC 307 are similar mass or more massive than the highest mass bin from \citetalias{Parikh2021}, which is 11 - 11.3 log(M/M$_{\odot}$) and shown in red.

Our ages and metallicities are larger (older and more metal rich) than the MaNGA-based results. We also find shallower gradients in the present analysis. Note that the total metallicity in this work was not derived directly, but calculated using [Fe/H] and other abundances. N, Na, and Mg are also consistent between the two analyses. NGC 5516 and NGC 5328 are outliers with lower [N/Fe] than the other galaxies of similar masses and dispersions by 0.2~dex. Meanwhile C, Ca, and Ti are offset to lower abundances for the current analysis. These differences are likely due to a combination of the choice of stellar population models and the wavelength range of the data. We cannot repeat our present analysis with the \cite{ThomasD2011b} as in \citetalias{Parikh2021} since the MUSE data does not cover the bluer wavelengths containing the necessary absorption features. A comprehensive discussion of the treatment of these elements in the different models and comparison of previous literature results (which show similar offsets as here) is included in \citet{Parikh2019}. It remains to be investigated what impact the differences between the abundances of these particular elements might have on the derived IMF slope.

\subsection{Comparison with dynamical M/L}
\label{sec:disc_ML}
\begin{figure}
  \includegraphics[width=\linewidth]{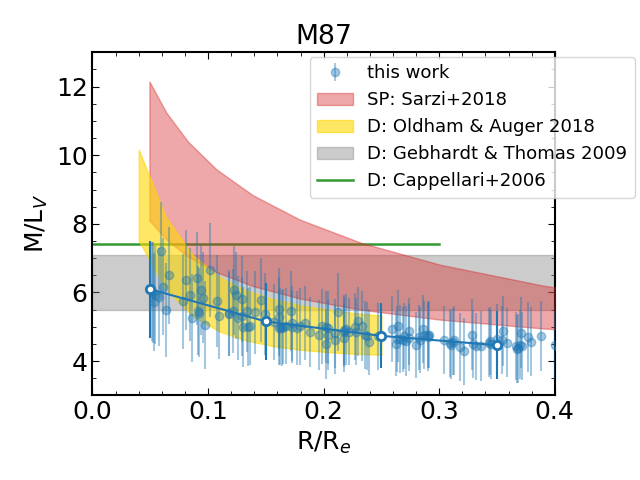}
\caption{Comparison of our M/L (blue symbols) with results from literature for M87, where \textit{SP} denotes stellar populations-based analyses and \textit{D} denotes dynamical analyses. Results from our Voronoi binned data are shown as filled blue circles, while the radial bins are shown as open circles connected by lines. \citet{Sarzi2018} values (red) from fitting indices using a different set of stellar population models are systematically larger than our results by $\sim$50\%. Our results have excellent agreement with the dynamical models of \citet{Oldham2018} (yellow). Results from other dynamical models with a constant M/L (\citet{Gebhardt2009}, grey; \citet{Cappellari2006}, green) have similar values, differences are due to aperture sizes and different modelling techniques. This also highlights the importance of including M/L gradients in dynamical models to allow consistent comparisons.}
\label{fig:M87_sps_dyn_ML}
\end{figure}

\begin{figure}
  \includegraphics[width=.9\linewidth]{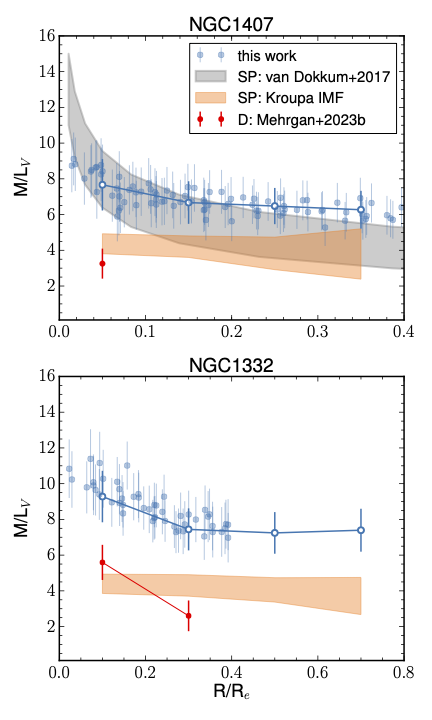}
\caption{Comparison of M/L for NGC 1407 and NGC 1332 with literature,
  where \textit{SP} denotes stellar populations-based analyses and
  \textit{D} denotes dynamical analyses. Our results are shown in blue
  for different binning schemes: Voronoi bins (filled circles), radial
  bins (open circles connected by lines). Our NGC 1407 results agree
  well in the centre with \citet{vanDokkum2017} which are based
  on different data but same method and models. For both galaxies,
    dynamical M/L are shown in red. We focus on spatial scales where dark-matter does not affect the stellar M/L
    which restricts the comparison to the innermost radial bin for most galaxies.
    In general, stellar population
    results and dynamics agree in M/L that increase towards the
    center. However, dynamical gradients are confined to sub-kpc
    scales \citep{Mehrgan2023b} and are not resolved by the binning
    scheme used here. The bin-averaged M/L from dynamics are closer to
    Kroupa (orange shaded regions) than stellar populations.
  }
\label{fig:sps_dyn_ML2}
\end{figure}

\begin{figure*}
  \includegraphics[width=.8\linewidth]{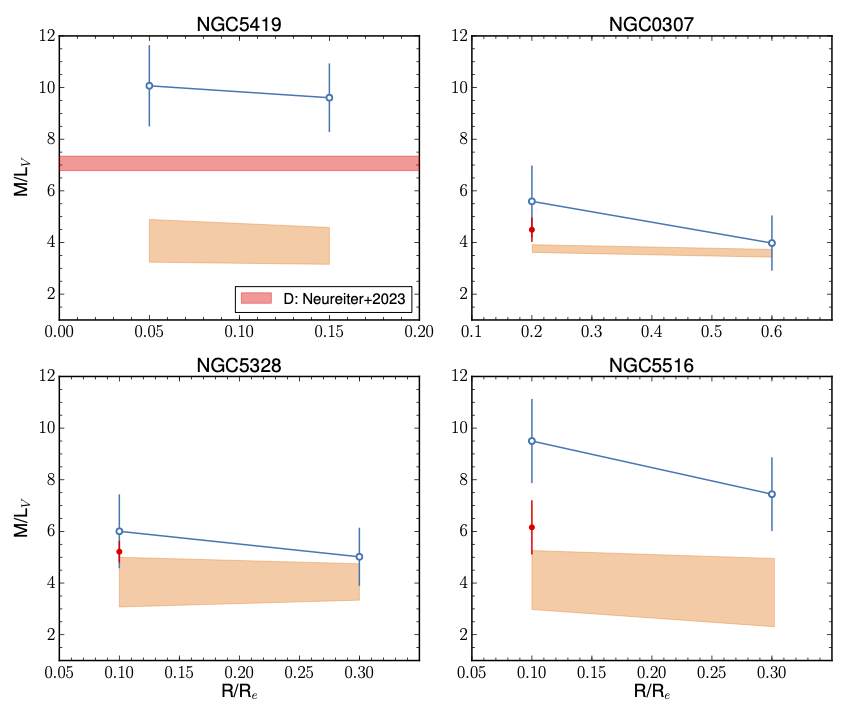}
\caption{Comparison of our M/L (blue) with dynamical
    models (red) for the other four
    galaxies. These have been analysed in radial bins, shown as open circles. M/L from dynamics are systematically lower
    than from stellar populations. Except for NGC5419 they are close
    to the prediction for a Kroupa IMF (shown as orange shaded regions).
}
\label{fig:sps_dyn_ML3}
\end{figure*}

We now present a comparison of the M/L derived in this work (always
shown in blue) with literature results from stellar populations and
dynamics-based analyses, where available. All M/L are reported in the
V band. We start with M87 in \autoref{fig:M87_sps_dyn_ML}, which has
been analysed a lot in literature.
NGC 1407 and NGC 1332 are shown in
\autoref{fig:sps_dyn_ML2} and the remaining four galaxies are shown in
\autoref{fig:sps_dyn_ML3}. NGC 7619, for which we analysed a single
global spectrum, is discussed in the text. Literature results are
labelled D for dynamics or SP for stellar populations. Shaded regions
and error bars represent 1-$\sigma$ errors. The shaded region from
\citet{Sarzi2018} includes systematics due to different methods used
to infer the IMF.

For all galaxies except M87, SINFONI kinematics with
  high spatial resolution are available and all these galaxies have
  advanced orbit-based dynamical models that include a central supermassive
  black hole, stars and a dark-matter halo \citep{Mehrgan2023b,Neureiter2023b}.
  The
  combination of MUSE and SINFONI together with a novel model
  selection technique allowed for highly accurate mass determinations
  \citep{Lipka2021,Thomas2022,deNicola2022,Neureiter2023a}. Moreover,
  \citet{Mehrgan2023b} probed for spatial variations of M/L and found
  very concentrated stellar M/L gradients on sub-kpc scales. We
  compare these new dynamical M/L inside the radius
  $r_\mathrm{main}$ (cf. \citealt{Mehrgan2023b}) where the mass
  decomposition between stellar and dark matter is very robust and
  average the dynamical gradients over the size of the radial bins
  used here.


\textit{M87:} Our results from Voronoi bins are shown as filled blue symbols in \autoref{fig:M87_sps_dyn_ML}, while open blue circles show the elliptical annuli. Stellar population results from \citet{Sarzi2018} (red) are based on a different set of models which could explain their larger M/L. The shaded region for \citet{Sarzi2018} includes systematics due to the method of IMF determination. They made a comparison with the dynamical models from \citet{Oldham2018} (yellow) and noted that the stellar population M/L were systematically larger. We find smaller M/L values, 5.5-7.5 in the centre and 4.5-5 at 0.4~R$_e$, eliminating this systematic shift from the dynamical models.

Dynamical models with constant M/L are also shown from \citet{Gebhardt2009} (grey region) and \citet{Cappellari2006} (green line). These results are based on Schwarzschild and Jeans Anisotropic Modelling respectively, as well as different aperture sizes, which could be the reason for the difference. The three analyses that allow for a varying M/L with radius find remarkably similar gradients, which is a very encouraging result.



\textit{NGC 1407 and NGC 1332:} Moving on to
\autoref{fig:sps_dyn_ML2}, we show our results for the other two
galaxies from Voronoi bins (filled circles), radial bins (open symbols
with lines). Our results for NGC 1407 agree with \citet{vanDokkum2017}
(red), but they find a steeper radial gradient. We use the same
stellar population models and fitting procedure but their data is from
the LRIS on the Keck I telescope, and includes the Wing-Ford band at
9920~\AA\ in the spectral range. However this region of the spectrum
is not fit very well and shows the largest residuals (see their figure
9). Their data also include C and Ca absorption features at bluer
wavelengths up to 4000~\AA, however we find only mild differences
(0.05~dex) between our abundance values.  Also shown
  are dynamical models from \citet{Mehrgan2023b}.


We find large central M/L for the fast rotator NGC 1332, and a
flattening in the outer parts like NGC 1407. Dynamics are systematically below
  stellar populations and closer to the Kroupa level (shown in orange).
  

\textit{NGC 5419, NGC 307, NGC 5328, NGC 5516:} We find M/L gradients
in these galaxies as shown by the results from the radial bins. The
stellar populations always give systematically larger M/L than dynamics. While \citet{Mehrgan2023b} also report
  M/L that increase towards the center, the scale of these gradients
  is mostly within our central bin and the average of dynamics
  is closer to Kroupa (see also the discussion in
  \citealt{Mehrgan2023b}). The result from triaxial dynamical
modelling by \citet{Neureiter2023b} is shown for NGC 5419 as the red shaded region.


\textit{NGC 7619:} For this galaxy we only analysed one average
spectrum out to 0.2~R$_e$, with M/L = $7.1\pm1.5$.
The models of \citet{Mehrgan2023b} yield 
  $5.3\pm1.4$ over the same spatial scale, again systematically lower.

During our analysis we explored various sources of systematic
errors. Possible systematics in the stellar population results could
arise due to the models, including incompleteness of the underlying
stellar library, and the assumed shape of the IMF. We discuss each of
these in the following sections. For the dynamical models,
uncertainties in M/L can arise from systematics in the dark matter
halo models or from assumptions about axisymmetry \citep{Thomas2007,
  Rusli2013, deNicola2022, Neureiter2023a}.
  
\begin{figure}
  \includegraphics[width=.9\linewidth]{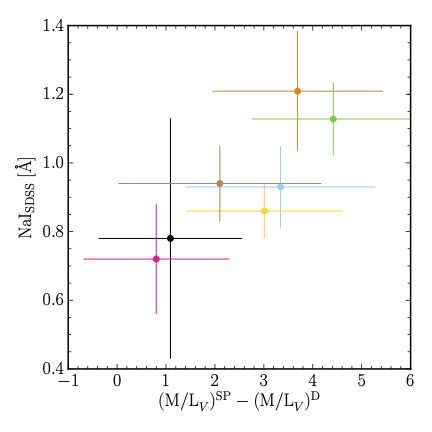}
  \caption{Difference between dynamics and stellar
      populations versus the NaI absorption strength. The stronger
      the NaI index, the more the stellar populations are biased high compared to dynamics, suggesting possible calibration issues of this line in combination with correlated uncertainties.
  }
\label{fig:sps_dyn_nai}
\end{figure}

\subsection{Models}

Updated stellar population models described in \citet{Maraston2020}
based on the MaStar stellar library \citep{Yan2018} vastly expand on
the available stellar parameter space. Low mass dwarf stars are not
well represented in other stellar libraries and are either included
using a theoretical approach or via extrapolation. The new stellar
library contains such stars very close to the hydrogen burning limit
for the first time. Figure 16 in \citet{Maraston2020} shows that for
stellar population models based on these stars, the NaI absorption is
much stronger in comparison to other widely used models. This would
lead to shallower IMFs and smaller M/L. 
  Fig.~\ref{fig:sps_dyn_nai} shows that the difference between
  dynamics and stellar populations indeed increases with NaI. The
  Spearman correlation coefficient for this trend is $r_s = 0.86$ with
  a significance level of $p=0.014$ indicating a strong
  correlatin. Calibration of the NaI absorption might therefore be a
  significant source of systematics between dynamics and stellar
  populations. A potential correlation with the abundance-sensitive
  NaD is weaker ($r_s = 0.71$ with $p=0.07$). We suggest that a combination of correlated errors and the calibration of the NaI absorption is causing the systematics between dynamics and stellar populations.

\subsection{IMF parameterisation}
\begin{figure*}
  \includegraphics[width=.85\linewidth]{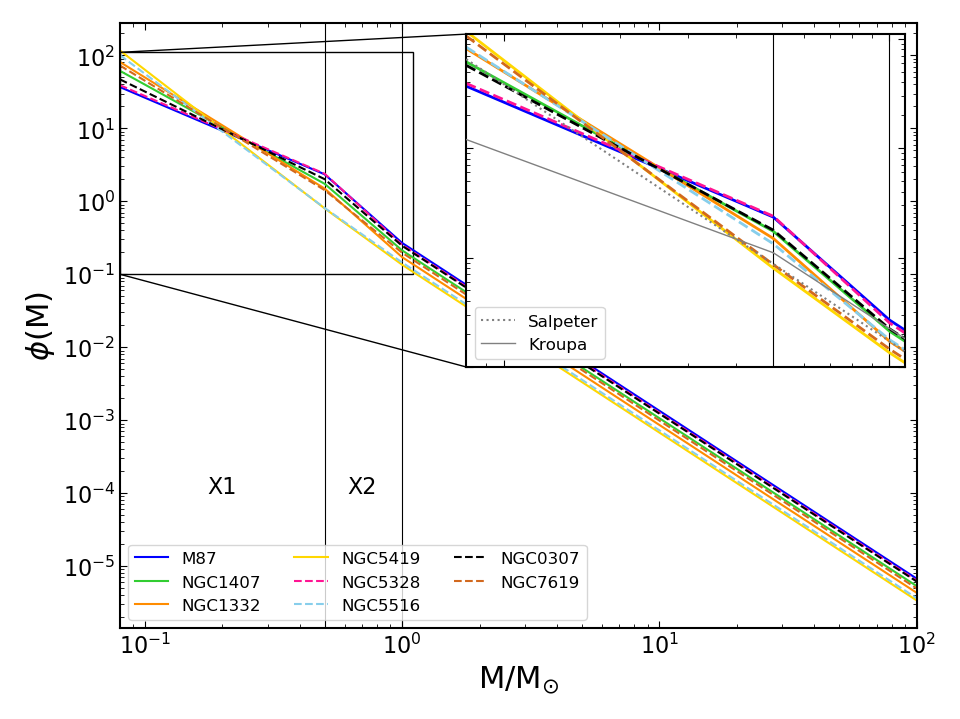}
\caption{The IMF shape as derived for our galaxy sample is shown. X1 and X2 are free parameters, while the slope above 1~M$_{\odot}$ is always fixed to 2.3. The inset zooms in on the low mass end, Kroupa and Salpeter IMFs are shown for reference. All IMFs are normalised to have the same mass. X1 clearly shows more variation among the galaxies.}
\label{fig:imfshape}
\end{figure*}

Results also depend on the assumed shape of the IMF. \autoref{fig:imfshape} shows the derived IMF slope as a function of stellar mass for our sample of galaxies. The physical reason why the IMF slope would change from 2.3 above 1~M$_{\odot}$ to values of 3 at intermediate masses, and falling again below a Salpeter slope at the lowest masses, is unclear.

We repeated our fits while fixing the IMF slope above 0.5~M$_{\odot}$, X2, to the Salpeter value, and keeping the slope between 0.08 - 0.5~M$_{\odot}$, X1, free. This parameterisation was also used in our previous work \citepalias{Parikh2018}. With X2 fixed to 2.3, we found that X1 increased to large values, leading to an overall increase in M/L. E.g. for NGC 1407 we derived M/L$_V\ =\ 9.8\pm1.3$ and $8.2\pm1.4$ for the inner and outermost bins respectively, larger than both the \citet{vanDokkum2017} results as well as the dynamics.

\citet{Newman2017} explore a varying low-mass cutoff and find better agreement when the cutoff is larger at 0.15~M$_{\odot}$, or by using non-parametric models with a turnover below 0.3~M$_{\odot}$. We do not have access to non-parametric models but it would be important to test what happens when making no assumptions about the IMF shape.

\section{Conclusions}
\label{sec:Conclusion}
We analysed the detailed stellar populations of a sample of eight massive early-type galaxies using MUSE data. Using full spectrum fitting we derived ages, metallicities, element abundances, IMF slopes, and M/L on high S/N binned spectra. Our conclusions are as follows:

\begin{enumerate}
\item The galaxies host old ages and negative metallicity gradients. They are all [Mg/Fe] enhanced to at least twice solar values, with M87 most enhanced at +0.4~dex. We also present abundances of C, N, Na, Ca, and Ti, and compare these with literature.
\item There is a range in IMF low-mass slope X1, with radial gradients, while X2 has high values $\sim$3 at all radii. This leads to excess mass compared to a Kroupa IMF, with $\alpha_{Kroupa}$ between 1.5 - 2.5 for galaxy centres. We find no trend with kinematic morphology, a larger sample is needed to explore this in detail.
\item M/L gradients from stellar populations and dynamics show
  excellent agreement for M87. We derive smaller M/L than stellar
  population results from literature based on different
  models. Meanwhile updated models with missing dwarf stars could lead
  to M/L smaller than our results, highlighting the uncertainty due to
  the stellar population synthesis and underlying stellar library.
\item For the other galaxies we compare to
  state-of-the-art orbit-based dynamical M/L with
    gradients and find stellar populations systematically higher than
    dynamics. The discrepancy increases systematically with the NaI
    absorption strength, suggesting that improved calibration of the
    NaI absorption could be the key to reconciling dynamics and stellar
    populations.
\item There is strong motivation for measuring dynamical stellar M/L gradients of individual galaxies as an independent comparison with stellar population-based results to test models and identify sources of systematics.
\end{enumerate}

\section*{Acknowledgements}
We thank Daniel Thomas for fruitful discussions. Computations were performed on the high performance computing (HPC) resources of the Max Planck Computing and Data Facility in Garching, and the Sciama HPC cluster which is supported by the Institute of Cosmology of Gravitation, SEPnet and the University of Portsmouth. TP acknowledges Development Projects Funding from the University of Portsmouth. This research made use of the \textsc{python} packages \textsc{numpy} \citep{VanDerWalt2011}, \textsc{scipy} \citep{Jones2001}, \textsc{matplotlib} \citep{Hunter2007}, and \textsc{astropy} \citep{Astropy2013}.

\section*{Data Availability}
Data generated by the analysis in this work are available upon reasonable request from the corresponding authors.



\bibliographystyle{mnras}
\bibliography{Refs} 



\begin{appendix}

\section{Fits}
\label{sec:app_fits}
The best-fitting models to the inner and outermost radially binned spectra are shown for the other galaxies in our analysis.

\begin{figure*}
  \includegraphics[width=.9\linewidth]{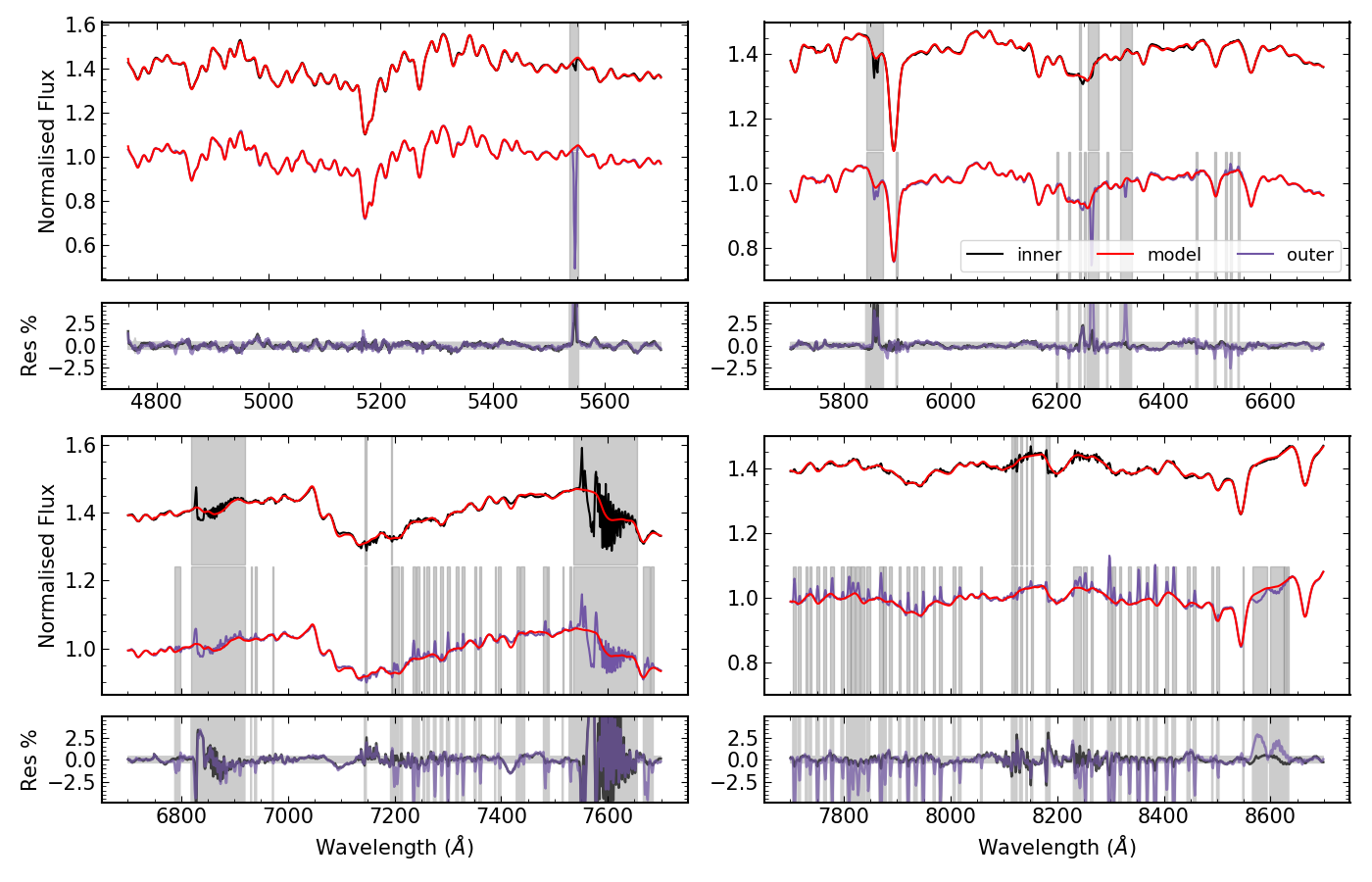}
  \caption{Stellar population model fit to the central and outermost radially binned spectrum of NGC 1407. Masked regions affected by sky lines and telluric absorption are shown in grey. Residuals between the data and model are shown in the bottom panels, and the horizontal grey band shows the error on the data.}
\label{fig:radial_fit_NGC 1407}
\end{figure*}

\begin{figure*}  
  \includegraphics[width=.9\linewidth]{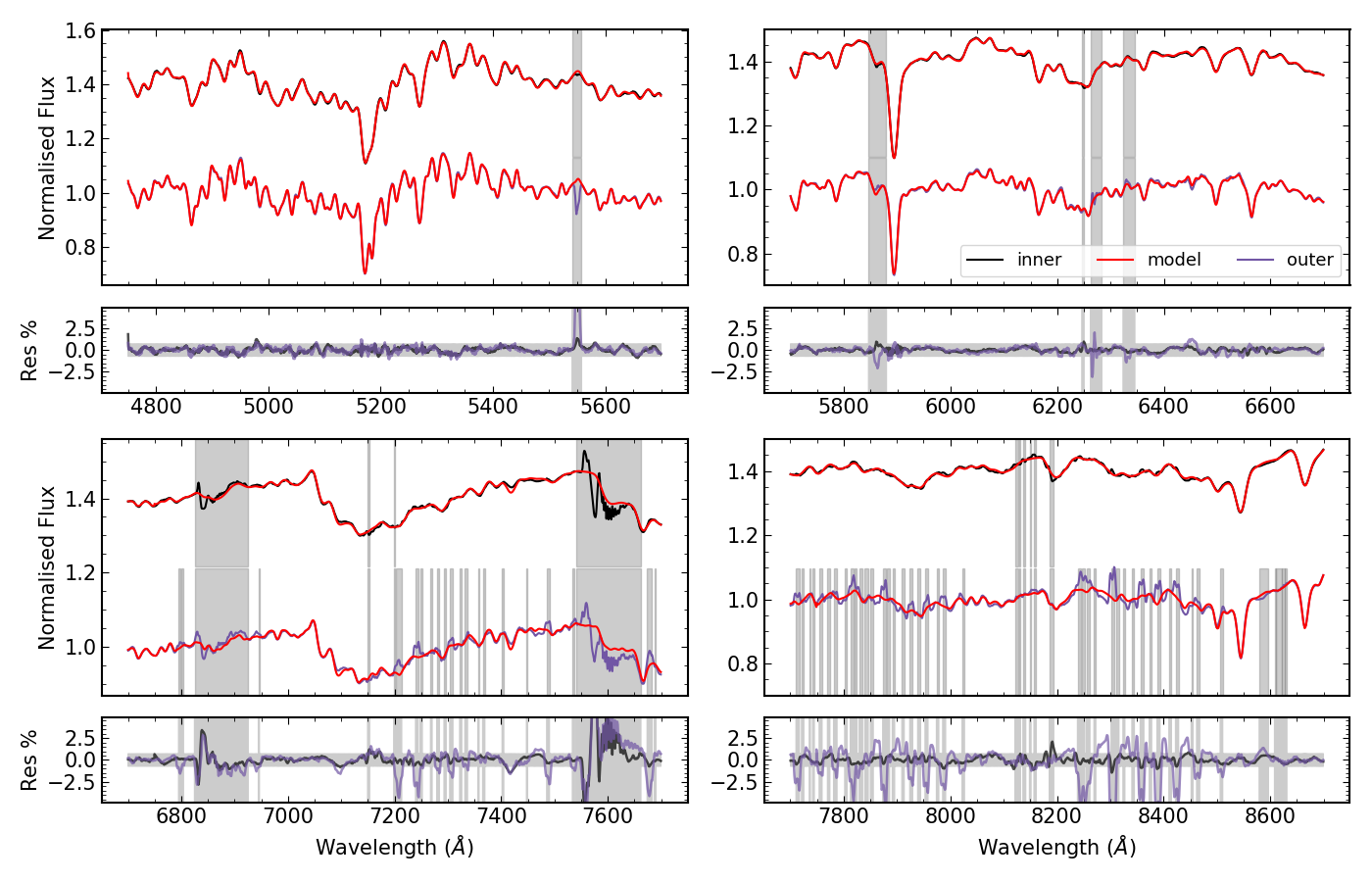}
  \caption{Same for NGC 1332.}
\label{fig:radial_fit_NGC 1332}
\end{figure*}

\begin{figure*}
  \includegraphics[width=.9\linewidth]{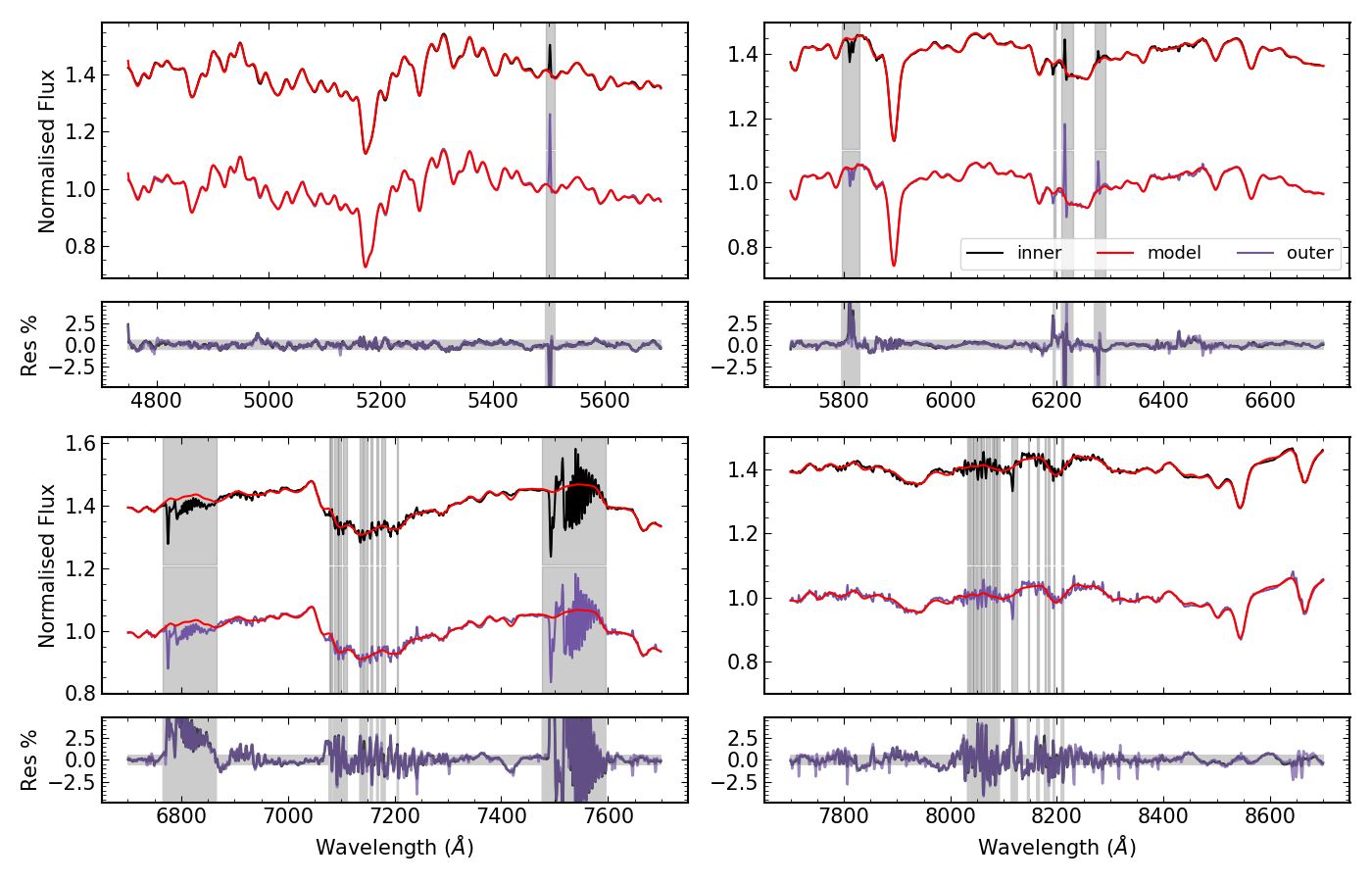}
\caption{Same for NGC 5419.}
\label{fig:radial_fit_NGC 5419}
\end{figure*}  

\begin{figure*}
  \includegraphics[width=.9\linewidth]{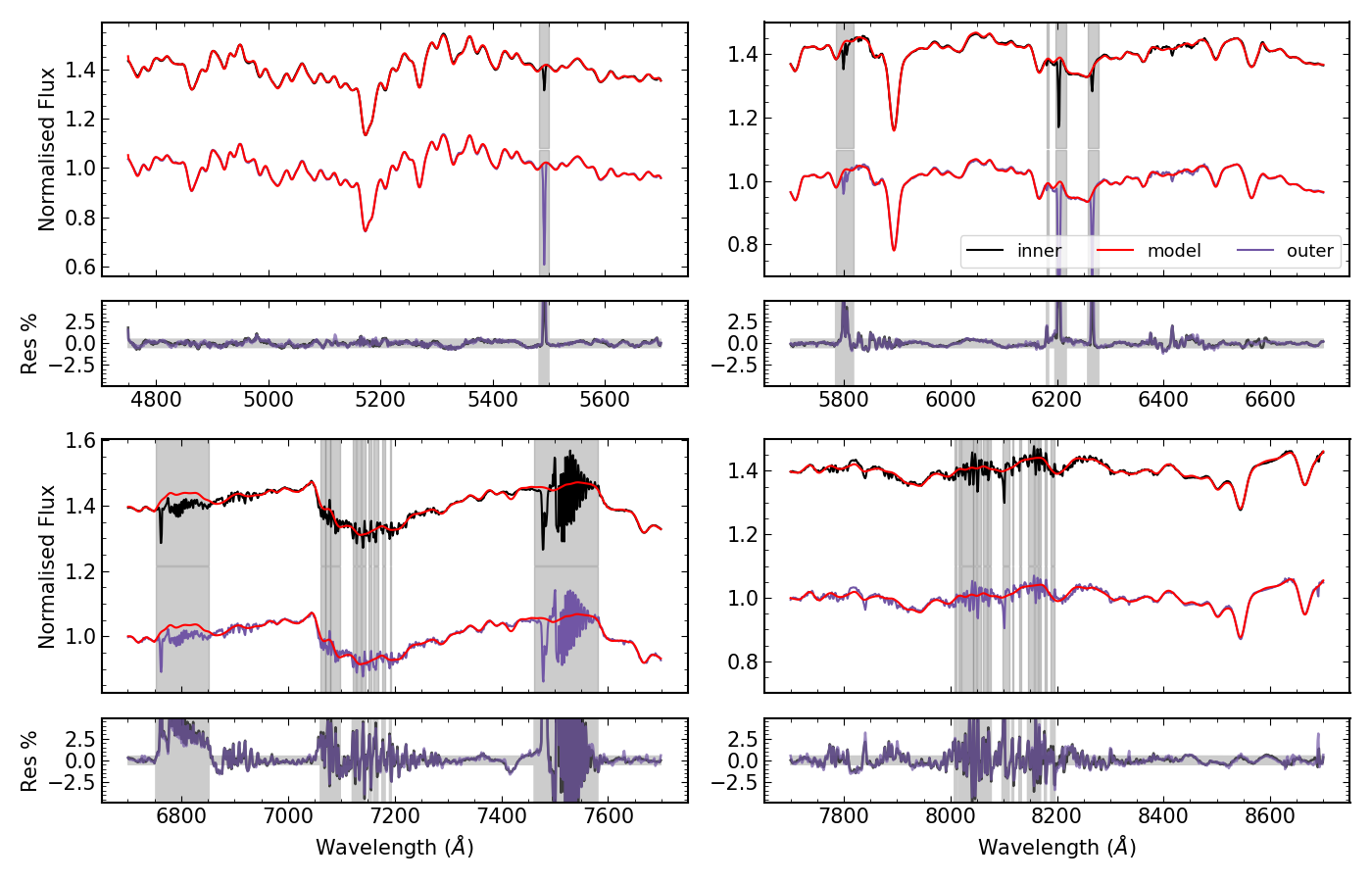}
  \caption{Same for NGC 5328.}
\label{fig:radial_fit_NGC 5328}
\end{figure*}

\begin{figure*}
  \includegraphics[width=.9\linewidth]{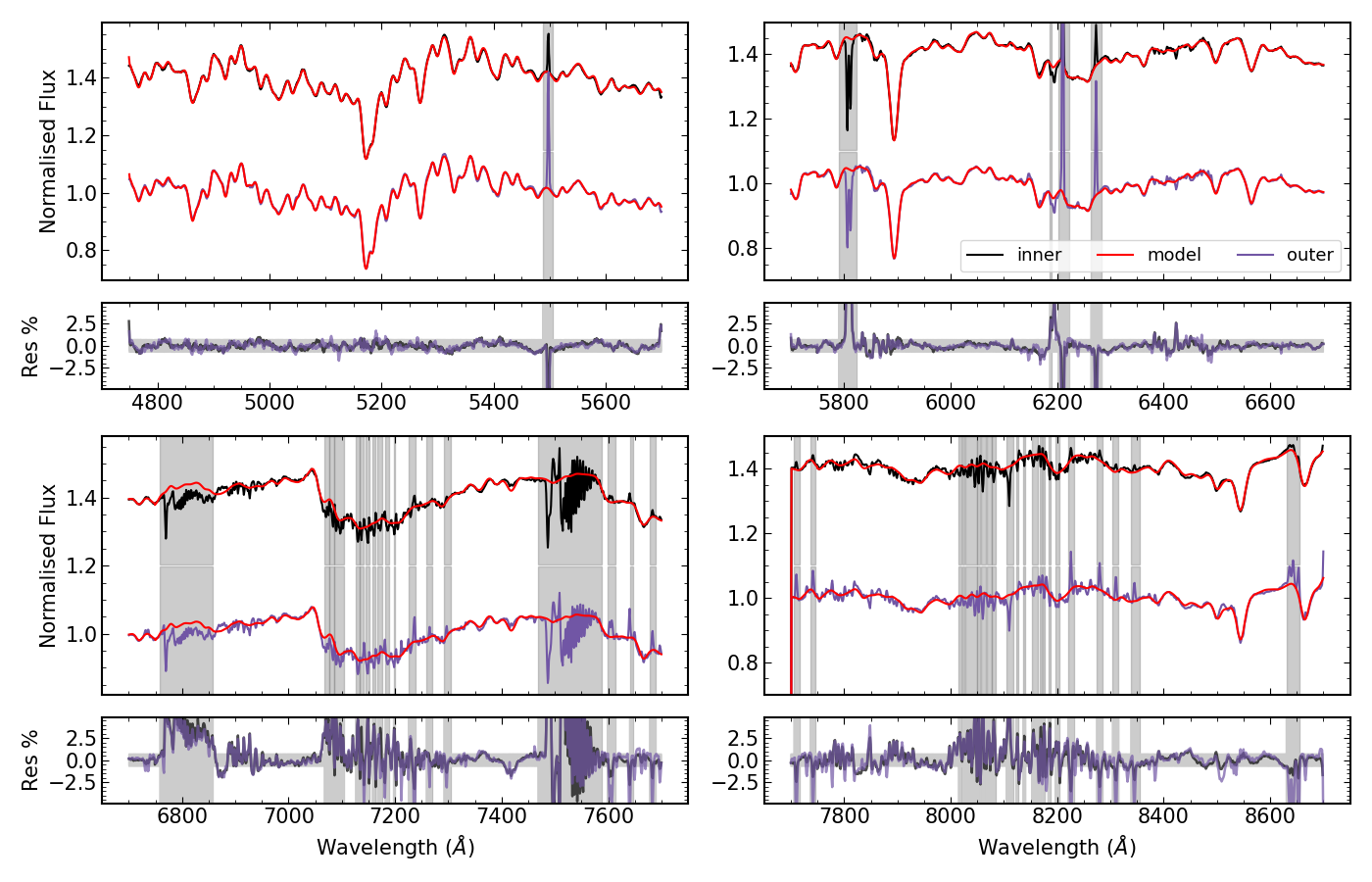}
  \caption{Same for NGC 5516.}
\label{fig:radial_fit_NGC 5516}
\end{figure*}

\begin{figure*}
  \includegraphics[width=.9\linewidth]{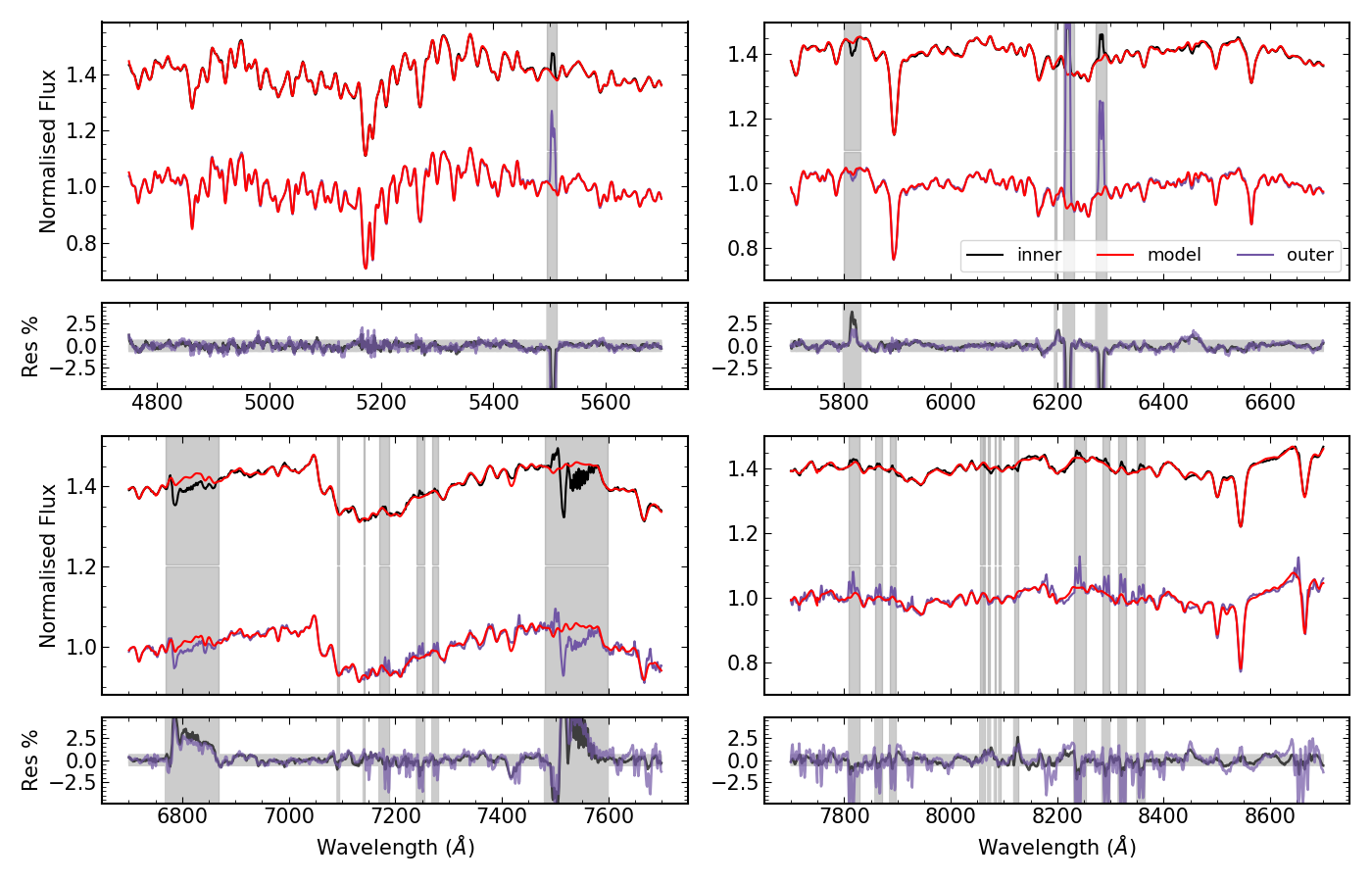}
  \caption{Same for NGC 307.}
\label{fig:radial_fit_NGC 0307}
\end{figure*}
  
\begin{figure*}
  \includegraphics[width=.9\linewidth]{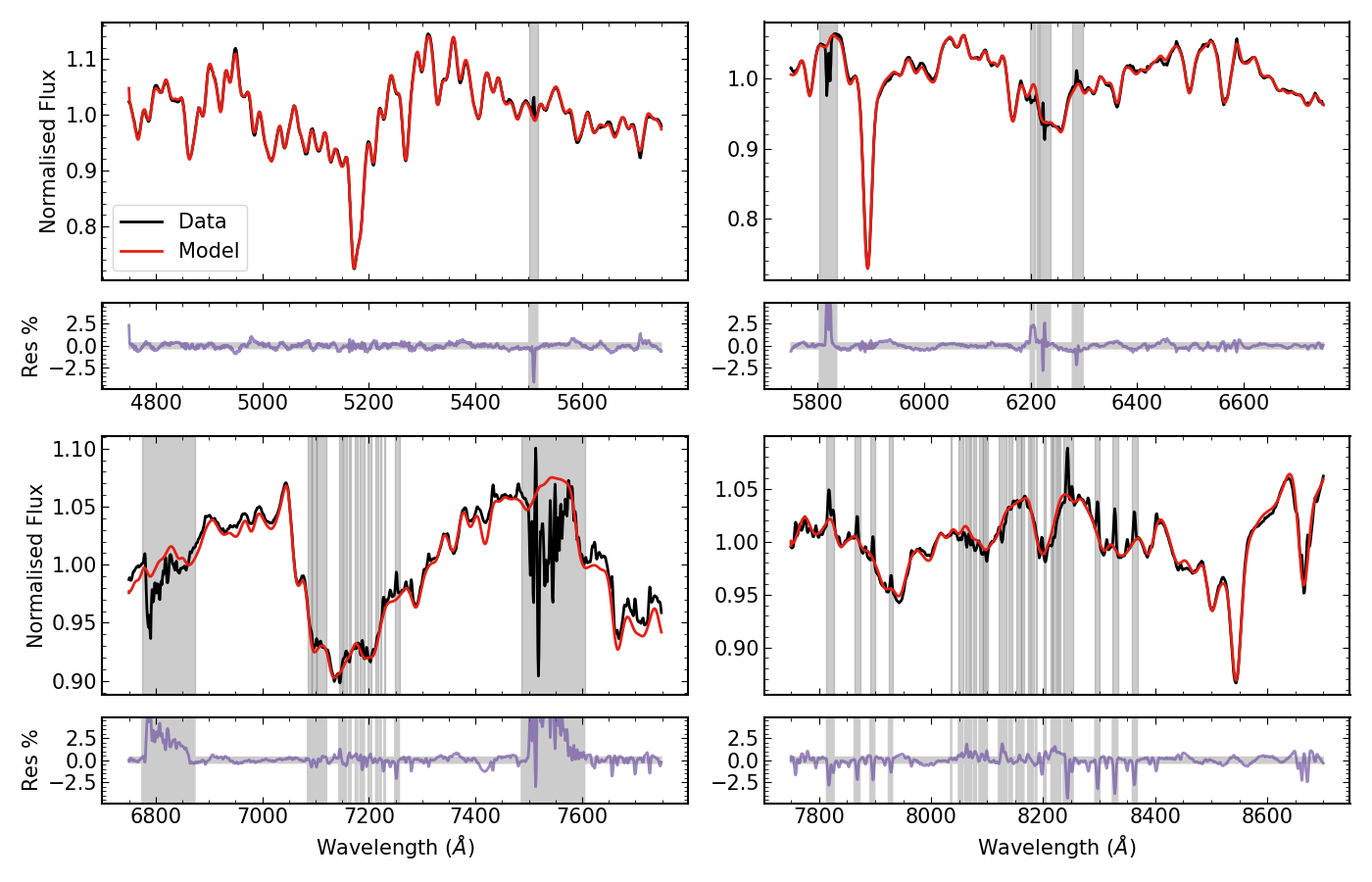}
  \caption{Same for NGC 7619.}
\label{fig:radial_fit_NGC 7619}
\end{figure*}

\section{Maps}
\label{sec:app_maps}
The stellar population parameters derived from full spectrum fitting are shown as maps for the remaining two galaxies in the detailed Voronoi binning scheme.

\begin{figure*}
  \includegraphics[width=.8\linewidth]{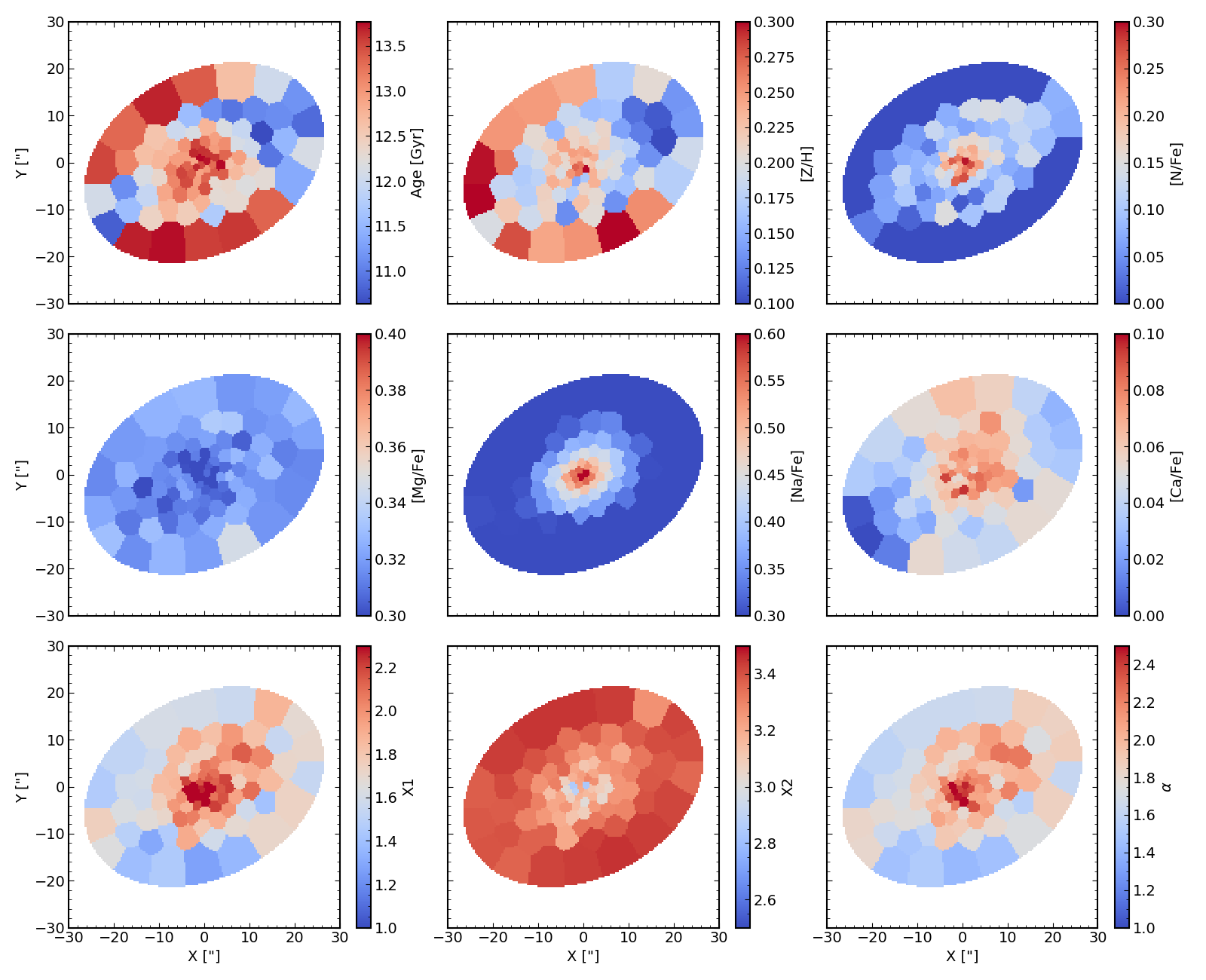}
\caption{Stellar population maps for NGC 1332.}
\label{fig:NGC 1332_maps}
\end{figure*}

\begin{figure*}
  \includegraphics[width=.8\linewidth]{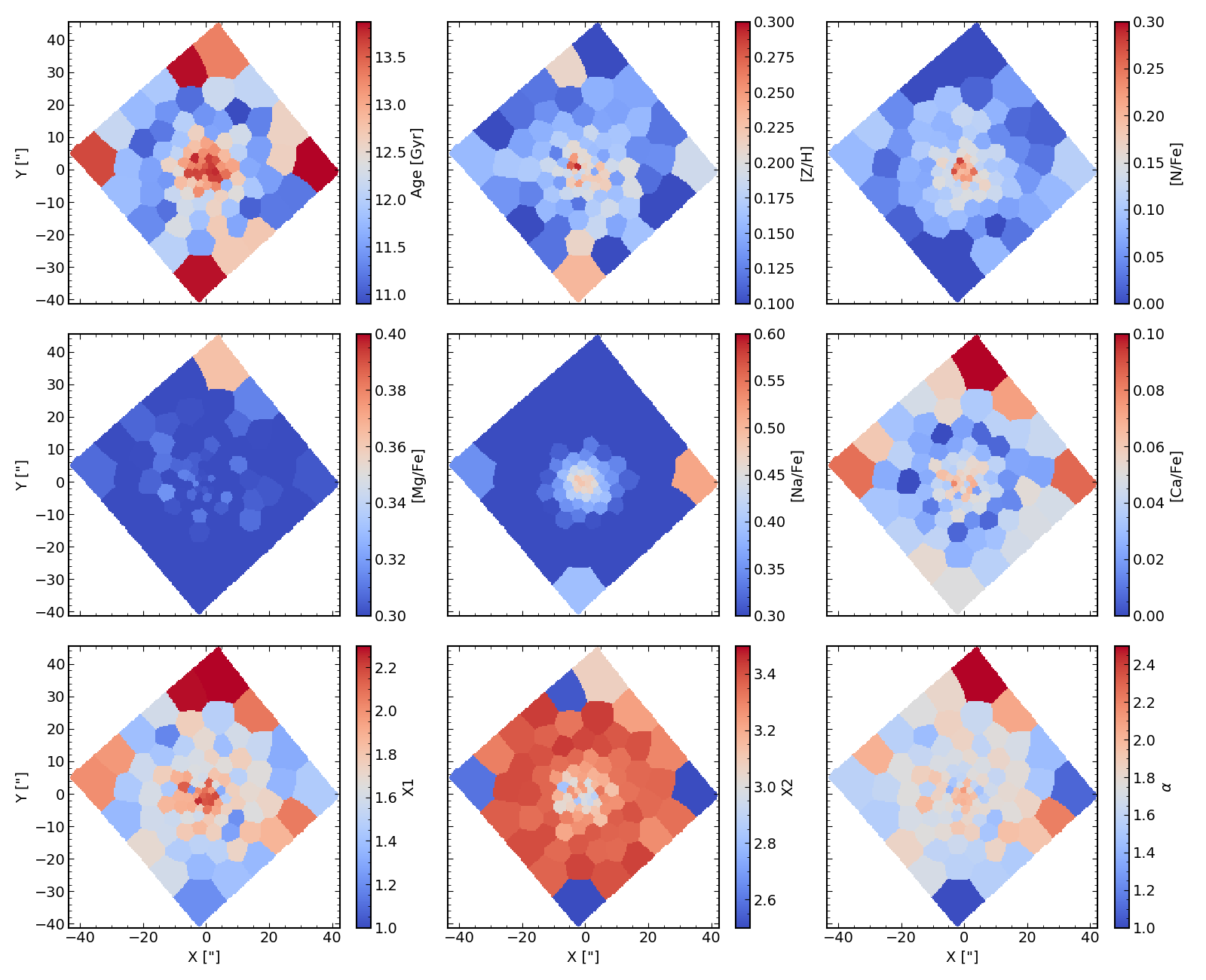}
\caption{Stellar population maps for NGC 1407.}
\label{fig:NGC 1407_maps}
\end{figure*}

\section{LOSVD shape}
\label{sec:app_h3h4}
\begin{figure*}
 	\includegraphics[width=.7\linewidth]{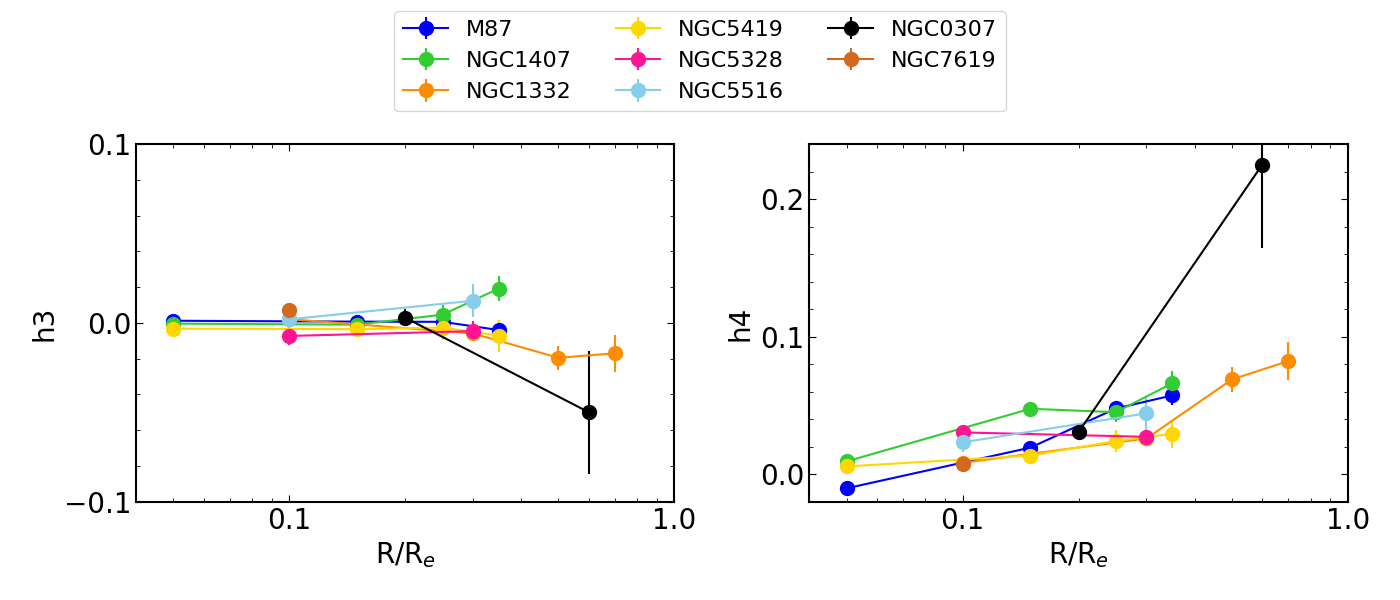}
    \caption{Higher order Gauss-Hermite moments, h3 and h4, as a function of radius for radially binned spectra of all galaxies. The LOSVD appears to be Gaussian in galaxy centres, while h4 rises at large radii for FRs (NGC 1332 and NGC 307), suggesting that the stellar population parameters are unaffected by biases due to the LOSVD shape apart from in these regions.}
    \label{fig:h3h4}
\end{figure*}

We also derive higher order Gauss-Hermite moments, h3 and h4, from the full spectrum fitting. For the Voronoi binned galaxies, we find maps fully consistent with \citet{Mehrgan2023a}, giving confidence in the derived parameters from ALF. Generally the results show that LOSVD wings are stronger in galaxy centres and for FRs. \autoref{fig:h3h4} shows these parameters for the radial bins. Small h3 and h4 are measured in galaxy centres, suggesting that these LOSVDs are Gaussian, and hence unaffected by biased parameters. However at large radii, for fast rotators we see sharp rises in h4, e.g. NGC 1332 and NGC 307. This is because the galaxies have strong rotation in the outskirts, and averaging these spectra together distorts the LOSVD shape. 

%

\end{appendix}


\bsp	
\label{lastpage}
\end{document}